\documentclass[aps,twocolumn,amsmath,amssymb,preprintnumbers,nofootinbib]{revtex4}
\usepackage{amsmath} \usepackage{amsfonts} \usepackage{amssymb}
\usepackage{bbm}
\usepackage{epsfig}
\usepackage{graphics}
\usepackage{graphicx}
\newcommand{\be}{\begin{equation}}
\newcommand{\ee}{\end{equation}}
\newcommand{\ba}{\begin{eqnarray}}
\newcommand{\ea}{\end{eqnarray}}
\newcommand{\nn}{\nonumber}

\newcommand{\A}{^{(A)}}

\newcommand{\kl}{\langle}
\newcommand{\kr}{\rangle}

\begin{document}

\title[ ]{Probabilistic observables, conditional correlations, and quantum physics}

\author{C. Wetterich}
\affiliation{Institut  f\"ur Theoretische Physik\\
Universit\"at Heidelberg\\
Philosophenweg 16, D-69120 Heidelberg}

\begin{abstract}
We discuss the classical statistics of isolated subsystems. Only a small part of the information contained in the classical probability distribution for the subsystem and its environment is available for the description of the isolated subsystem. The ``coarse graining of the information'' to micro-states implies probabilistic observables. For two-level probabilistic observables only a probability for finding the values one or minus one can be given for any micro-state, while such observables could be realized as classical observables with sharp values on a substate level. For a continuous family of micro-states parameterized by a sphere all the quantum mechanical laws for a two-state system follow under the assumption that the purity of the ensemble is conserved by the time evolution. The correlation functions of quantum mechanics correspond to the use of conditional correlation functions in classical statistics. We further discuss the classical statistical realization of entanglement within a system corresponding to four-state quantum mechanics. We conclude that quantum mechanics can be derived from a classical statistical setting with infinitely many micro-states.

\end{abstract}

\maketitle

\section{Introduction}
\label{introduction}
Quantum statistics is often believed to be fundamentally different from classical statistics. In quantum statistics, the complex probability amplitudes and transition amplitudes play a key role. Probabilities only obtain as squares of the amplitude, and this gives rise to spectacular phenomena as interference and entanglement. In contrast, classical statistics is directly formulated in terms of positive probabilities. Furthermore, unitarity is the most characteristic feature of the time evolution in quantum mechanics. This aspect is not easily visible in the time evolution of classical probabilities. Finally, the quantum mechanical uncertainty principle is based on the non-commutativity of the operator product, while the “pointwise” product of observables in classical statistics is obviously commutative.

We argue that the difference between quantum statistics and classical statistics is only apparent. We demonstrate that quantum mechanics can be described as a classical statistical system with infinitely many states. In this paper we mainly concentrate on a system that is equivalent to a two-state quantum system. We consider discrete observables that can only take values $\pm 1$. They correspond to the spin operators of the equivalent quantum system. We will obtain the characteristic features of non-commuting spin operators in a classical setting. All the usual uncertainty relations of quantum mechanics are directly implemented. We also generalize our setting to a classical ensemble which is equivalent to four-state quantum mechanics. This allows us to describe the classical statistical realization of entanglement and interference.

We formulate the condition for the time evolution of our simplest classical ensemble that leads to the unitary transformations characteristic for the quantum evolution. It involves the concept of purity of a statistical ensemble. A purity conserving time evolution in classical statistics is equivalent to the unitary time evolution in quantum mechanics. Pure classical states are those where one of the discrete observables takes a sharp value, say $+1$. This means that the probability vanishes for all states where the value of the observable takes a value different from one. Pure classical states correspond to pure quantum states and can be described by a wave function. We derive the von Neumann and Schr\"odinger equations for these states. 

We take here the attitude that the basic description of reality should be probabilistic, while an (almost) deterministic behavior arises only in limiting cases. In particular, we do not attempt a deterministic local hidden variable theory. However, even on the level of a probabilistic theory it is widely believed that quantum mechanics needs statistical concepts beyond classical statistics, while we argue here that the classical statistical concepts are sufficient and quantum mechanics can emerge from a classical statistical probability distribution. 

In particular a classical statistical setting admits the definition of different conditional correlation functions for the description of the outcome of sequences of measurements. This takes into account that measurements can change the state of the system, and that this change may depend on the type of measurement. For a two-state system only one particular conditional correlation function allows predictions which use only the information available within the two-state system, without invoking information from the environment. This conditional correlation differs from the classical correlation which is based on joint probabilities.

The reader may cast strong doubts about these statements from the beginning. The big conceptual puzzles of quantum mechanics, as the Einstein-Rosen-Podolski paradoxon \cite{EPR}, have triggered a lot of attempts to replace quantum mechanics by a more fundamental deterministic theory. Based on Bell’s inequalities \cite{Bell} for correlators of entangled states it was argued that such attempts cannot succeed, since quantum correlations contradict either realism or locality. We will argue in the last section that both locality and a version of ``probabilistic realism'', where the elements of reality can be described by correlation functions as well as values of observables, can be maintained. However, the classical statistical systems which describe quantum systems miss another property that is usually implicitly assumed in the derivation of Bell's inequality, namely the property of ``completeness'' of the statistical system. Here a statistical system is called complete if joint probabilities for the values of all pairs of observables are defined and if the ``measurement correlation'' for pairs of observables is determined by those joint probabilities. We argue that the subsystems which show quantum mechanical properties are described by ``incomplete statistics'' \cite{3} in this sense. This is closely related to a ``coarse graining of the information'' if one concentrates on properties only involving the subsystem. 

It can be shown \cite{CHSH}, \cite{B71}, \cite{CH}, \cite{CS} that Bell’s inequalities apply if  two measurements are appropriately described by the classical correlation function $\langle A\cdot B\rangle$ for two observables $A$ and $B$. The ``classical'' or ``pointwise'' correlation function $\langle A\cdot B\rangle$ means that the observables $A$ and $B$ have both fixed values in a state of the ensemble, which are multiplied and averaged over the ensemble. Such systems are statistically complete. Deterministic local ``hidden variable theories'' usually assume this property. While Bell's inequalities indeed exclude such deterministic local hidden variable theories, they are not necessarily in contradiction with a classical statistical formulation of quantum mechanics which employs conditional correlations.

\section{Outline}
\label{Outline}
Our classical statistical implementation of two-state and four-state quantum mechanics is based on four main ingredients: (i) The quantum system is described by an isolated subsystem of a classical statistical ensemble with infinitely many degrees of freedom. It can be characterized by a restricted set of probabilistic observables. (ii) The state of the subsystem is determined by the expectation values of a set of ``basis observables''. (iii) Conditional correlations, which can be computed from the state of the subsystem, describe the outcome of sequences of measurements. (iv) The unitary time evolution of the subsystem is implemented by particular properties of the time evolution of the probability distribution of the classical ensemble.

The concept that is perhaps least familiar is the use of conditional correlations in a classical statistical ensemble. Indeed, within classical statistics one can define correlation functions different from the classical or pointwise correlation. We advocate that the outcome of two measurements of the observables $A$ and $B$ should not be described by the pointwise correlation $\langle A\cdot B\rangle$, but rather by conditional correlations $\langle A\circ B\rangle$, which are related to a different product structure $A\circ B$. This can best be understood for two subsequent measurements. In general, the first measurement ``changes the state of the system'' by eliminating all possible sequences of events which contradict this measurement. The second measurement is performed with new conditions, depending on the outcome of the first measurement. The idealized situation where the effect of the first measurement on the state of the system can be neglected may be realized for some large systems as classical thermodynamics, but not for systems with only a few effective degrees of freedom, as often characteristic for those described by quantum theory. If the outcome of the first measurement matters for the second, conditional probabilities should be used. Conditional correlations require a specification of the state of the system after the first measurement. The underlying elimination of possibilities contradicting the first measurement is not unique, however. We argue that for proper measurements of properties of the (sub-) system only information available for the system should be employed for the specification of the state after the first measurement. The details of the state of the environment should not matter. This excludes the use of the classical correlation function, except for certain special or limiting cases. 

We define within classical statistics a ``conditional product'' $A\circ B$ of two observables, and the associated ``conditional correlations''. They only involve information available for the subsystem. We show how to express the conditional correlation functions in terms of quantum mechanical operator products. The conditional correlations for classical probabilistic observables equal the appropriate conditional correlations defined in quantum mechanics. If the correct conditional correlations are used for a description of two consecutive measurements, we obtain the same results for the classical statistics and the quantum description. No conflict with Bell’s inequalities arises for the classical statistics implementation of quantum mechanics. We propose that the conditional correlation or ``quantum correlation'' should be used for the general description of two measurements in classical statistics, and not only if two measurements are clearly separated in time. The perhaps more familiar classical correlation arises from the quantum correlation only for appropriate limiting cases. 

We do not consider the present work as only a formal or mathematical reformulation of quantum mechanics. We have a rather physical picture in mind, where the classical statistical setting describes an atom simultaneously with its environment. This is analogous to the role of an atom in quantum field theory, where it appears as a particular excitation of a highly complicated vacuum - its ``environment''. Quantum mechanical properties can arise when the statistical description focuses on the atom and discards all information pertaining to the environment. Typically, the state of the atom is described by only a few quantities. In the simplest case of an atom with spin one half in the ground state, the (sub-) system will be described by only three real numbers $\rho_k$ (neglecting the motion of the atom in space and excited energy levels). One therefore is interested in possible observables - and structures among them - that can be described in terms of the reduced information contained in $\rho_k$, rather than involving the full information contained in the classical probability distribution which describes both the system and its environment.

The embedding of the quantum mechanical concepts within a more general classical statistical setting, which also includes the environment, permits us to ask new questions. What are the particular conditions for a quantum mechanical description to hold? Why do we observe all small enough systems in nature as quantum systems? We advocate that the unitary time evolution in quantum mechanics expresses the ``isolation'' of the subsystem from its environment. (This does not mean that the environment can simply be omitted from the classical description of the subsystem.) Interactions with the environment can lead to the phenomena of decoherence or ``syncoherence'' - the approach of a mixed state to a pure state. 

In this paper we present a rather detailed account of the classical statistics description of a quantum mechanical two-state system. We investigate discrete observables which can only take the values $\pm 1$. A useful notion is the concept of ``probabilistic observables'' which are characterized by a probability to find the value $+1$ or $-1$ in every micro-state, rather than by a fixed value in a given micro-state. Such a generalized notion of ``fuzzy observables'' \cite{POB}, \cite{BB} is well known in measurement theory \cite{Neu}.

Probabilistic observables can be implemented as classical observables if the micro-state consists of several or even infinitely many substates. In other words, once part of the degrees of freedom of a classical statistical system - the substates - are ``integrated out'', a classical observable on the substate level becomes a probabilistic observable on the level of the remaining micro-states. The quantum mechanical pure and mixed states will be associated with particular micro-states. A typical observable may have a sharp value for particular micro-states, but typically a probability distribution of different values in the other micro-states. 

The general notion of a probabilistic observable is a rather wide concept. One has to specify the amount of information available, in particular concerning joint probabilities for pairs of two probabilistic observables. In our approach this is adapted to the picture of a subsystem within a classical statistical ``environment''. On the level of the micro-states two observables $A$ and $B$ are, in general, not comeasurable. This means that the joint probability $p_{ab}$ of finding the value $a$ for $A$ and $b$ for $B$ is not defined.  \footnote{In this context it may be interesting to  compare our setting for probability distributions and observables to the so called ``classical extension'' of quantum statistical systems \cite{BB}, \cite{SB}. This mathematical approach shows certain similarities, but also important differences to our setting. For a ``classical extension'' one constructs a probability distribution for classical states which correspond to the pure quantum states \cite{MIS}, and one extends the notion of observables such that two observables which correspond to non-commuting operators in quantum mechanics become comeasurable. (As an example, the extension provides the joint probability for two spin components $S_x$ and $S_z$ to have both the value up or $+1$. Such a joint probability is not available in quantum mechanics.) In our approach, comeasurability is not given on the level of micro-states, even though we may construct on this level probability distributions on the manifold of pure quantum states. Instead, we can realize comeasurability on the level of the classical substates which describe the system and its environment. However, only a submanifold of the possible probability distributions for the substates can be mapped to the  quantum states. (It may nevertheless be possible that one can formally construct a ``classical extension'' also in our approach - so far this does not play a role for the description of the system.)}

The notion of micro-states is introduced in this paper partly for purposes of gaining intuition. It demonstrates in a simple way how the notion of probabilistic or fuzzy observables, which is characteristic for observables in a quantum state, arises naturally in a classical statistical setting. Alternatively, our approach could proceed directly from classical states with fixed values of classical observables (the substates in this paper) to the quantum system. This procedure is followed in refs. \cite{CWN,14A}, where the mathematical structures underlying our concepts are discussed in more detail. In ref. \cite{14A} we also give simple explicit realizations of the classical statistical ensembles which describe the quantum system and its environment. In a formulation based on the substates the notion of micro-states needs not to be introduced and one can proceed directly to the quantum states of the subsystem. 

The setting is then simply classical statistics with all classical observables taking fixed values in all states. Nevertheless the notion of micro-states and probabilistic observables may be considered as a useful way to organize the statistical information for certain specific systems. The mapping from the classical observables on the substate level to the probabilistic observables on the micro-state level is not invertible. We will see that the operators in quantum mechanics correspond to the probabilistic observables. Due to the lack of invertibility no map which associates to each quantum operator a classical observable exists. For this reason the Kochen-Specker-theorem \cite{KS} does not apply. On the other hand, the implementation of probabilistic observables as classical observables on the level of substates is actually not necessary. One may, alternatively, treat the probabilistic observables as genuine objects of a classical statistical description of reality. In this version the Kochen-Specker theorem finds no application because the classical observables do not have fixed values. A more detailed discussion of the properties of observables and states can be found in \cite{CWN}, \cite{14A}.

Beyond a statistical setting for states and observables other key features of quantum mechanics have to be implemented in a classical statistical setting. This concerns, first of all, a prescription for predictions of the outcome of two (or more) measurements - an issue related to the concept of correlation and discussed extensively in this paper. Many physicists believe that a proper probabilistic setting for quantum mechanical observables is not the central distinction between classical statistics and quantum mechanics, but rather the issue of correlations. We share this opinion. Furthermore, one has to understand the quantum mechanical time evolution, starting from the time evolution of a classical probability distribution. This requires an explicit construction of the density matrix in terms of the classical probability distribution. At the end, the issue will be to understand how quantum mechanical behavior emerges for physical systems within a more general classical statistical setting. 

The classical statistical description of quantum mechanics can be generalized to systems with more than two quantum states. In a four-state system we have described the phenomenon of entanglement between two two-state subsystems \cite{N}. Entanglement is often believed to be the center piece of quantum statistics. It is a central issue of many theoretical discussions about the foundations of quantum mechanics, as decoherence \cite{DC} or the measurement process \cite{Zu}, and it underlies the idea of quantum computing\cite{Zo}. Spectacular experiments on teleportation \cite{Ze} rely on it. A classical statistics description of entanglement may find useful practical applications and influence the conceptual and philosophical discussion based on this phenomenon. We give a short account of four-state quantum mechanics in the later part of this paper. There is no limitation in the number of quantum states $M$. Taking $M\to\infty$ yields continuous quantum mechanics. In particular, the quantum particle in a potential has been described by a classical statistical ensemble in this way \cite{19A}.

This paper is organized as follows. In sect. III we discuss the notion of probabilistic observables, with particular emphasis on two-level observables that may also be called spins. Sect. IV compares the realization of rotations in the classical and quantum statistical setting. The classical system needs an infinity of micro-states if a continuous rotation is to be realized. At least one ``classical pure state'' must exist for every rotation-angle. In sect. V we reduce the infinity of classical micro-states to a finite number of ``effective states''. All expectation values of the spin observables can be computed in the effective state description in terms of three numbers $\rho_k$. The prize of the reduction is, however, that the ``effective probabilities'' $\rho_k$ are not necessarily positive anymore. In sect. VI we construct the density matrix $\rho$  of quantum mechanics from the effective probabilities. We establish for our classical statistical ensemble the quantum mechanical rule for the computation of expectation values of observables, $\langle A\rangle=tr(A\rho)$.

Sect. VI turns to the issue of correlation functions and introduces the conditional product of two observables and the conditional correlation.  The conditional two point function is commutative. This does not hold for the higher conditional correlation functions - the three point function is not commutative since the order of consecutive measurements matters. Sect. VII completes the mapping between classical statistics and quantum statistics. We introduce the wave function for pure states and relate conditional probabilities to squares of quantum mechanical transition amplitudes. We then derive the expression of the conditional correlations in terms of quantum mechanical operator products. The non-commutativity of the conditional three point function can be directly traced to non-vanishing commutators of operators. The derivation of these results demonstrates the use of quantum mechanical transition amplitudes for questions arising in a classical statistical setting. A simple example for a classical ensemble with a finite number of degrees of freedom is presented in sect. \ref{simpleexample}. It describes three cartesian spins without a continuous rotation symmetry.

In sect. \ref{timeevolution} we deal with the time evolution. We define the purity $P$ of a statistical ensemble - in our simplest case 
\begin{equation}\label{0}
P=\sum_k\rho^2_k=2 \text{tr} \rho^2-1. 
\end{equation}
A purity conserving time evolution amounts to the unitary time transformation of quantum mechanics. Furthermore, a more general time evolution of the classical ensemble can describe decoherence for decreasing purity, as well as ``syncoherence'' for increasing purity. We argue that the quantum mechanical pure states correspond to partial fixed points of the more general time evolution in classical statistics. In sect. \ref{pseudoquantum} we briefly discuss classical systems with a finite number of states $N$, which may be used to obtain quantum mechanics in the limit $N\to\infty$. Sect. \ref{realizations} discusses the possible realizations of probabilistic observables. We generalize in sect. \ref{four} our construction to classical ensembles that correspond to four-state quantum mechanics. We show the classical realization of entanglement, interference and the distinction between bosons and fermions. Finally, we summarize in sect. \ref{probabilisticrealism} the conceptual issues of realism, locality and completeness for statistical systems and discuss Bell's inequalities. Our conclusions are presented in sect. \ref{conclusions}. 

\section{Probabilistic observables}
\label{probabilisticobservables}

In this section we discuss the basic notion of probabilistic observables which do not have a sharp value in a given micro-state.

\medskip\noindent
{\bf  1.\quad Expectation values}

Consider a probabilistic system with $N$ classical micro-states, labeled by $\sigma =1...N$, and characterized by probabilities $p_\sigma \geq0~,~\sum\limits_\sigma  p_\sigma =1$. A classical or deterministic observable $A^{(cl)}$ is specified by $N$ real numbers $\bar A_\sigma $, such that the expectation value reads
\begin{equation}\label{1}
\langle A\rangle=\sum^N_{\sigma =1}\bar A_{\sigma }p_\sigma .
\end{equation}
In a given micro-state $\sigma $ the classical observable has a {\em fixed} value, namely $\bar A_\sigma $. The probabilistic nature of the system arises only from the probabilities to find a given micro-state $\sigma $. 

This concept can be generalized by introducing probabilistic observables, for which we can only give probabilities to find a certain value in a given micro-state $\sigma $. Probabilistic or fuzzy observables are well known in measurement theory and have been investigated for quantum and classical systems \cite{POB}, \cite{BB}. We give here a simple description of the properties relevant for our discussion. A probabilistic observable is characterized by a set of real functions $w_\sigma (x)\geq 0$, normalized according to $\int dx w_\sigma (x)=1$. The expectation values of powers of the probabilistic observables in a {\em given} micro-state $\sigma $ obey
\begin{equation}\label{2}
A^Q_\sigma =\int dx x^Qw_\sigma (x)~,~\bar A_\sigma =\int dxxw_\sigma (x).
\end{equation}
Correspondingly, the expectation values in a macro-state of the probabilistic system reads
\begin{equation}\label{3}
\langle A^Q\rangle=\sum_\sigma  A^Q_\sigma  p_\sigma ~,~\langle A\rangle=\sum_\sigma \bar A_\sigma  p_\sigma .
\end{equation}

Classical observables correspond to the special case
\begin{equation}\label{4}
w_\sigma (x)=\delta(x-\bar A_\sigma )~,~A^Q_\sigma =(\bar A_\sigma )^Q.
\end{equation}
In this case all moments $A^Q_\sigma $ are fixed in terms of the mean value in the micro-state $\sigma $, i.e. the moment for $Q=1~,~\bar A_\sigma =\int dxxw_\sigma (x)$. In contrast, for the most general probabilistic observables the infinite set of moments $A^Q_\sigma $ may be used in order to parameterize the distribution $w_\sigma (x)$. For the most general probabilistic observable much more information is therefore needed for its precise specification, namely infinitely many real numbers $A^Q_\sigma $ instead of the $N$ real numbers $\bar A_\sigma $ for a classical observable. The probabilistic nature of the system is now twofold. It arises from the probability distribution $w_\sigma (x)$ to find the value $x$ of the observable in the micro-state $\sigma $, and from the probability distribution for the micro-states, $\{p_\sigma \}$, characterizing a given macro-state or ensemble. The relation \eqref{1} for the expectation value of $A$ remains valid for probabilistic observables. However, the expectation values of higher powers  $A^Q~,~Q\geq 2$, as given by eq. \eqref{3}, may differ from classical observables (cf. eq. \eqref{4}).

\medskip\noindent
{\bf  2. \quad Two-level observables}

As a specific example for a probabilistic observable we concentrate in this paper on the bi-modal distribution
\ba\label{5}
&&w_\sigma (x)=\frac12(1+\bar A_\sigma )\delta(x-1)+\frac12 (1-\bar A_\sigma )\delta(x+1),\nonumber\\
&&-1\leq\bar A_\sigma \leq 1~,~\bar A_\sigma=\int dxxw_\sigma (x),\nonumber\\
&&A^2_\sigma =\int dxx^2w_\sigma (x)=1.
\ea
In any micro-state $\sigma $ the observable can only take the values $+1$ or $-1$. In other words, for a given micro-state $\sigma $ the observable is specified by the relative probabilities to find the values $+1$ or $-1$, 
\begin{equation}\label{GA}
w_{\sigma  +}=(1+\bar A_\sigma)/2~,~w_{\sigma _-}=(1-\bar A_\sigma )/2. 
\end{equation}
Thus $N$ real numbers $\bar A_\sigma$ are again sufficient to specify the ``two-level observables'' obeying eq. \eqref{5}. The moments are given by 
\begin{equation}\label{6}
A^Q_\sigma =\left\{\begin{array}{llll}
\bar A_\sigma&\text{for}&Q&\text{odd}\\1&\text{for}&Q&\text{even}
\end{array}
\right.,
\end{equation}
implying for the macro-state
\begin{equation}\label{7}
\langle A^Q\rangle=\left\{\begin{array}{cll}
\sum_\sigma \bar A_\sigma  p_\sigma &\text{for}&Q\text{ odd}\\
1&\text{for}&Q\text{ even}
\end{array}\right. .
\end{equation}

We may realize the ensemble or the macro-state by an infinite set of measurements with identical conditions. Each measurement realizes a particular microstate, and the $p_\sigma $ give the relative numbers how often a given micro-state $\sigma $ is encountered in the ensemble. Since for any given micro-state the two-level observable can only take the values $+1$ or $-1$, the series of measurements of $A$ will produce a series of values $+1$ or $-1$, with relative probabilities $w_\pm=\frac12(1\pm \langle A\rangle)$. This is an easy way to understand why $\langle A^2\rangle=1$ for arbitrary $\{p_\sigma \}$. The situation amounts exactly to a quantum mechanical spin $1/2$-system, with an appropriate normalization of the spin operator, say in the $z$-direction, $\hat s_z=(\hbar/2)\hat S_z$: each measurement will give one of the eigenvalues $\pm 1$ of the operator $\hat S_z$. We will see that the association of the probabilistic two-level observable $A$ with a quantum-mechanical spin can be pushed much further than the possible outcome of a series of measurements. We will therefore often denote the two-level observables by ``spins'', but the reader should keep in mind that we treat here with purely classical probabilistic objects. 

Two-level observables are the simplest non-classical probabilistic observables. By simple shifts they can be easily generalized to any situation where an observable can only take two values (two ``levels'') in any given micro-state, like occupied / empty. One bit is enough for the possible values of the observable in a micro-state $\sigma $, say $0$ for $x=-1$ and $1$ for $x=1$. Nevertheless, the specification of the probabilistic observable needs the real numbers $\bar A_\sigma$. Instead of a continuous distribution $w_\sigma (x)$ we can replace eq. \eqref{2} by a discrete sum 
\begin{equation}\label{8}
A^Q_\sigma =\frac12\sum_{x=\pm 1}x^Q(1+x\bar A_\sigma).
\end{equation}

\medskip\noindent
{\bf  3.\quad Substates}

A single two-level observable can be represented as a classical observable in an extended statistical system consisting of substates. As an example, we may associate all points within the circle in Fig. 1 with substates. (We may consider a finite resolution with a finite number of points or we can consider the limit where the number of points goes to infinity.)
\begin{figure}[h!tb]
\centering
\includegraphics[scale=0.3]{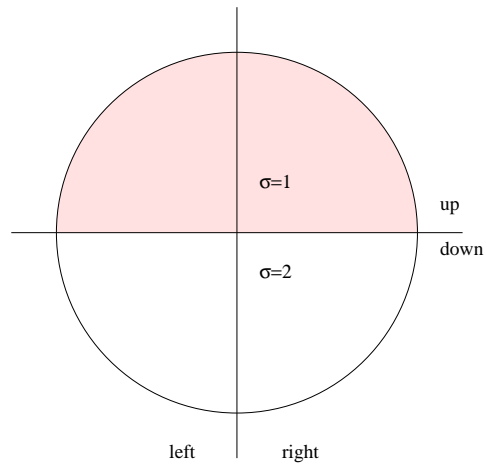}
\caption{Micro-states and two-level observables}
\label{fig2}
\end{figure}
\noindent
A characteristic two-level observable answers the question if the points are in the upper half plane or in the lower half plane. On the substate level the classical two-level observable $A^{(1)}$ takes the value $1$ for all points above the horizontal axis, and $-1$ for the points below the horizontal axis. If we denote the substates by $\tau$ one has in every state a fixed value, $A^{(1)}_\tau=\pm 1$. 

We next consider two micro-states $\sigma=1,2$. The first state $(\sigma=1)$ corresponds to a coarse graining where all points in the upper half plane are grouped together (shaded region in Fig. 1), while the second micro-state $(\sigma=2)$ combines the points in the lower half plane. The particular two-level observable $A^{(1)}$ remains a classical observable, with sharp values in the micro-states $\bar A^{(1)}_{\sigma=1}=1~,~\bar A^{(1)}_{\sigma=2}=-1$. We may also consider a second two-level observable $A^{(2)}$ for the decision between left and right. On the substate level it is again a classical observable, with $A^{(2)}_\tau=1$ for all points on the right of the vertical axis, and $A^{(2)}_\tau=-1$ if the substates are represented by points on the left side of the vertical axis. On the level of the two micro-states, however, the observable $A^{(2)}$ has no longer sharp values. For $\sigma=1$ (shaded region in Fig. 1) the micro-state groups together points from both the left and the right half. Typically, the observable $A^{(2)}$ has a distribution of values $\pm 1$ in this state, with $|\bar A^{(2)}_{(\sigma=1)}|<1$. Thus $A^{(2)}$ has to be described by a probabilistic observable on the level of micro-states.

We may label the substates $\tau$ according to the microstate $\sigma$ to which they ``belong'', $\tau=(\sigma,t_\sigma)$. (Here $t_\sigma$ distinguishes the states that belong to a given micro-state $\sigma$. For example, $t_{\sigma=1}$ are suitable coordinates for the points in the upper half plane in Fig. 1.) If the probability distribution $p_\tau$ on the substate level is known, the probabilities for the micro-states obey
\begin{equation}\label{10A}
p_\sigma=\sum_{t_\sigma}p(\sigma,t_\sigma).
\end{equation}
The mean value of the observable $A$ in the micro-state $\sigma$ is given by
\begin{equation}\label{10B}
\bar A_\sigma=\sum_{t_\sigma}A(\sigma,t_\sigma)p(\sigma,t_\sigma)/p_\sigma,
\end{equation}
where $p(\sigma,t_\sigma)/p_\sigma$ specifies the relative probabilities of two substates $t_\sigma$ for a given $\sigma$. It is easy to verify that the ensemble average of $A$ can be computed both on the substate or micro-state levels
\ba\label{10C}
\langle A\rangle&=&\sum_\tau p_\tau A_\tau=\sum_\sigma\sum_{t_\sigma}p(\sigma,t_\sigma)A(\sigma,t_\sigma)
\nonumber\\
&=&\sum_\sigma p_\sigma\bar A_\sigma.
\ea
For the observable $A^{(2)}$ we may compute the probability to find $A^{(2)}=1$ in a given micro-state $\sigma$ as 
\begin{equation}\label{10D}
w^{(2)}_{\sigma+}=\sum_{s_{\sigma+}}p(\sigma,+,s_{\sigma+})/p_\sigma
\end{equation}
where we have further decomposed the substates $t_\sigma$ as $t_\sigma=(\sigma',s_{\sigma\sigma'})~,~\sigma'=(+,-)$. (The states with $\sigma=1~,~\sigma'=+$ are represented by the points in the upper right quarter inside the circle of Fig. 1.) Unless $p(\sigma,-,s_{\sigma_-})=0$ for all substates $(\sigma,-,s_{\sigma-})$ one has $w^{(2)}_{\sigma+}<1$. Similarly, one finds $w^{(2)}_{\sigma+}>0$ unless $p(\sigma,+,s_{\sigma+})=0$ for all corresponding substates. The probabilistic observable $A^{(2)}$ becomes deterministic  in the micro-state $\sigma$ (sharp value) only if $w^{(2)}_{\sigma+}=1$ or $w^{(2)}_{\sigma+}=0$. With
\ba\label{10E}
&&w^{(2)}_{\sigma+}+w^{(2)}_{\sigma-}=1,\nonumber\\
\bar A^{(2)}_\sigma&=&\Big[\sum_{s_{\sigma+}}p(\sigma,+,s_{\sigma+})-
\sum_{s_{\sigma-}}p(\sigma,-,s_{\sigma-})\Big]/p_\sigma\nonumber\\
&=&w^{(2)}_{\sigma+}-w^{(2)}_{\sigma-}
\ea
we find consistency with eq. \eqref{GA}. 

In the coarse graining step from substates to micro-states most of the information contained in the probability distribution $p_\tau$ for the substates is lost. Instead of (infinitely) many numbers $p_\tau$ only two numbers, $w^{(1)}_+$ and $w^{(2)}_+$, are necessary to characterize the expectation values of the observables $A^{(1)}$ and $A^{(2)}$ and powers thereof. We will see later the correspondence between quantum states and micro-states - in both types of states observables have genuinely a distribution of values rather than fixed values. The state of a two-state quantum system will be fully characterized by the expectation values of three basis observables $A^{(k)}$. 

The coarse graining from the substate level to the micro-states should be interpreted in an abstract sense rather than being associated to resolution in space. In a general sense, a quantum system can be regarded as an ``isolated system'' within its environment. For example, we may regard an atom as an excitation of the vacuum, similar to the conceptual setting of quantum field theory. The vacuum is a complicated system, involving infinitely many degrees of freedom, which may be associated to the substates $\tau$. In contrast, an atom, say in the ground state which admits only two spin polarizations, involves only a few degrees of freedom. These degrees of freedom can be associated with an ``isolated system'' and will be represented on the level of micro-states with probabilistic observables.

As long as we consider only a single two-level probabilistic observable (say $A^{(2)}$) a minimal implementation as a classical deterministic observable does not require a large number of substates $\tau$. It is sufficient to assume that each micro-state $\sigma $ consists of two substates $\sigma +$ and $\sigma -$, for which the observable has either the sharp value $+1$ (for $\sigma _+$) or $-1$ (for $\sigma _-$). The probabilities for the substates are then given by $p_{\sigma  +}=p_\sigma (1+\bar A_\sigma )/2~,~p_{\sigma _-}=p_\sigma (1-\bar A_\sigma )/2$. We emphasize that a representation of probabilistic observables as classical observables on a substate level is possible, but not necessary. We could also consider probabilistic observables as a basic, more general definition of observables and never refer to substates. This will be discussed in sect. \ref{realizations}.

\medskip\noindent
{\em 4.\quad Operations among probabilistic observables}

Consider a probabilistic observable $A$ with a discrete and finite spectrum of possible measurement values $\gamma_a$. (The generalizations to a continuous or infinite spectrum are straightforward.) Besides the spectrum $\{\gamma_a\}$ a probabilistic observable is characterized by the associated probabilities $w_a(\sigma)$ for every micro-state $\sigma$. We can always define the multiplication with a constant $c$ by $A\to cA:~\gamma_a\to c\gamma_a$. Similarly, we may ``shift'' the probabilistic observable by adding a piece $s$ proportional to the unit observable, $A\to A+s:\gamma_a\to \gamma_a+s$. More generally, we can define $F(A)$ by $\gamma_a\to F(\gamma_a)$. However, the sum of two probabilistic observables $A,B$ is, in general, not defined. For different probability distributions $w^{(A)}_a(\sigma)~,~w^{(B)}_b(\sigma)$ it makes no sense to ``add the possible measurement values''. For a similar reason the product of $A$ and $B$ is not defined for general probabilistic observables.

A special situation arises if the joint probability to find simultaneously a value $\gamma^{(A)}_a$ for $A$ and $\gamma^{(B)}_b$ for $B$ is known for every micro-state. (We occasionally use a simplified notation where $\gamma\A_a$ is replaced by $a$.) We denote this probability by $w_{ab}(\sigma)$ and observe $w_a(\sigma)=\sum_b w_{ab}(\sigma)~,~w_b(\sigma)=\sum_a w_{ab}(\sigma)$. In this case of ``comeasurability'' one may use the generalized definition
\begin{equation}\label{9A}
\bar A_\sigma=\sum_{ab}\gamma_aw_{ab}(\sigma),
\end{equation}
and similar for $(\overline{f(A)})_\sigma$ or $(\overline{f(B)})_\sigma$. We can now define $A+B$ by adding $\gamma_a+\gamma_b$, and $AB$ by multiplying $\gamma_a\gamma_b$, 
\ba\label{9B}
(\overline{A+B})_\sigma&=&\sum_{ab}(\gamma_a+\gamma_b)w_{ab}(\sigma),
\nonumber\\
(\overline {AB})_\sigma&=&\sum_{ab}\gamma_a\gamma_bw_{ab}(\sigma).
\ea
This can be extended to general functions $F(A,B)$ according to 
\begin{equation}\label{9C}
( \overline{F(A,B)})_\sigma=\sum_{ab}F(\gamma_a,\gamma_b)w_{ab}(\sigma).
\end{equation}
We will see later that this special situation of comeasurable observables corresponds to commuting observables in quantum mechanics. On the other hand, the non-availability of a joint probability for general probabilistic observables will motivate us to describe sequences of measurements in terms of conditional correlations that do not involve the joint probabilities. This will turn out to be a basic ingredient for the violation of Bell's inequalities in quantum measurements.

For the example in Fig. 1 the spectrum $\gamma_a$ of the observable $A^{(1)}$ consists only of two values $\pm 1$, and similar for $\gamma_b$ for $A^{(2)}$. The joint probability $w_{++}(\sigma)$ for finding the values $A^{(1)}=1,A^{(2)}=1$ can be defined on the substate level as
\begin{equation}\label{13A}
w_{++}(\sigma=1)=w^{(2)}_{\sigma+}~,~w_{++}(\sigma=2)=0,
\end{equation}
with $w^{(2)}_{\sigma+}$ given by eq. \eqref{10D}. (For a minimal set of substates there is only one state $(\sigma,+)$ and therefore no sum of $s_{\sigma+}$ needed, $w^{(2)}_{\sigma+}=p(\sigma,+)/p_\sigma$.) The joint probability is available if one knows for each quarter of the circle the probability to be realized, i.e. if
\ba\label{13B}
p_{\sigma+}&=&\sum_{s_{\sigma+}}p(\sigma,+,s_{\sigma+})=w^{(2)}_{\sigma+}p_\sigma,\nonumber\\
p_{\sigma-}&=&\sum_{s_{\sigma-}}p(\sigma,-,s_{\sigma-})=w^{(2)}_{\sigma-}p_\sigma
\ea
are known. With
\begin{equation}\label{13C}
p_{\sigma+}+p_{\sigma-}=p_\sigma~,~\sum^2_{\sigma=1}p_\sigma=1
\end{equation}
this requires the knowledge of three independent numbers. On the level of micro-states and probabilistic observables these three numbers are, in general, not available, since the state of the system may be characterized by only two numbers, $\langle A^{(1)}\rangle$ and $\langle A^{(2)}\rangle$. 

We will discuss in sect. \ref{conditional} that the absence of knowledge of joint probabilities is a crucial aspect for the definition of correlations. The substate-probabilities \eqref{13B} involve properties of the quantum system together with its environment. They are not accessible by measurements involving only the quantum system. In sect. \ref{four} we will see the connection between the availability of information about joint probabilities in a quantum system and commuting quantum mechanical operators. For the systems investigated in sects. \ref{spinrotations}-\ref{realizations} the combined probability $w_{ab}(\sigma)$ for two independent probabilistic observables will not be available. This will be different in sect. \ref{four} where we consider observables that correspond to commuting quantum operators. For such ``commuting observables'' the joint probabilities are available. However, not all observables of the four-state quantum system discussed in sect. \ref{four} are mutually commuting. In sect. \ref{probabilisticrealism} we argue that the missing joint probabilities are the key for the understanding of characteristic quantum mechanical features as the violation of Bell's inequalities. 

\section{Spin rotations in classical statistics}
\label{spinrotations}

We may start with a single two-level observable or spin and a system with only two micro-states, i.e. the states $(+)(\sigma =1)$ and $(-)(\sigma =2)$, with $\bar A_1=1~,~\bar A_2=-1$. The expectation value reads $\langle A\rangle=p_1-p_2$. Since for all $\sigma $ one has $|\bar A_\sigma|=1$, this special case corresponds actually to a classical observable. The distribution \eqref{5} involves only one $\delta$-function, $w_1(x)=\delta(x-1)~,~w_2(x)=\delta(x+1)$. If we assign instead the values $\bar A_1=\bar A_2=0$ we encounter a genuinely probabilistic variable, leading in this case to a random distribution of $+1$ and $-1$ measurements. Our example in Fig. 1 corresponds to this simplest case with two micro-states. The first (classical) two-level observable corresponds to $A^{(1)}$, the second (probabilistic) observable to $A^{(2)}$. 

\begin{figure}[h!tb]
\centering
\includegraphics[scale=0.8]{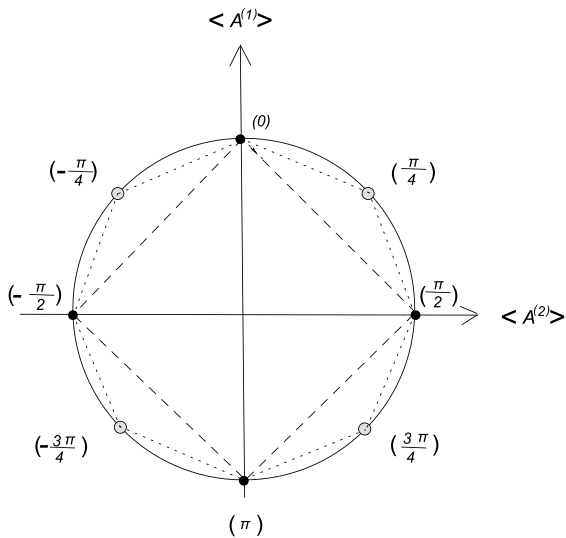}
\caption{Location of micro-states and expectation values of spins}
\label{fig1}
\end{figure}

An interesting case with two spin observables involves four classical micro-states, that we denote by $(+_1)$ or $(\pi)(\sigma =1)~,~(-_1)$ or $(-\pi)(\sigma =2)~,~(+_2)$ or $\left(\frac\pi 2\right)(\sigma =3)$ and $(-_2)$ or $\left(-\frac\pi 2\right)(\sigma =4)$, according to the full dots in Fig.~\ref{fig1}. 
The corresponding values of $\bar A^{(1)}$ and $\bar A^{(2)}$ are shown in the left half of Table \ref{table1}. 
\begin{table}[h!tb]
\begin{tabular}{|c|cccc|cccc|}
\hline
~&$(0)$&$(\pi)$&$\left(\frac\pi 2\right)$&$\left(-\frac\pi 2\right)$&
$\left(\frac\pi 4\right)$&$\left(-\frac\pi 4\right)$&$\left(\frac{3\pi}{4}\right)$&$\left(-\frac{3\pi}{4}\right)$\\ \hline
$\bar A^{(1)}_\sigma $&$1$&$-1$&$0$&$0$&
$\frac{1}{\sqrt{2}}$&$\frac{1}{\sqrt{2}}$&$-\frac{1}{\sqrt{2}}$&$-\frac{1}{\sqrt{2}}$\\
$\bar A^{(2)}_\sigma $&$0$&$0$&$1$&$-1$&$\frac{1}{\sqrt{2}}$&$-\frac{1}{\sqrt{2}}$&
$\frac{1}{\sqrt{2}}$&$-\frac{1}{\sqrt{2}}$\\ \hline
\end{tabular}
\caption{Mean values of spin observables in different micro-states}
\label{table1}
\end{table}
The expectation values of the two spins obey
\begin{equation}\label{9}
\langle A^{(1)}\rangle=p_1-p_2~,~\langle A^{(2)}\rangle =p_3-p_4.
\end{equation}
We note that both $A^{(1)}$ and $A^{(2)}$ are truly probabilistic observables since in some micro-states they have a zero mean value and thus equal probability for $+1$ and $-1$ values. We could also include observables with opposite mean values. Since they are obviously just a multiplication of $A^{(k)}$ by $-1$, we will not discuss them separately. 

On this level the probabilistic observables could again be expressed in terms of classical observables for a system with a higher number of states. We may assume that each micro-state consists of four substates with fixed values for $A^{(1)}$ and $A^{(2)}$, i.e. $++,+-,-+$ and $--$, such that we have a total of sixteen classical substates. Their probabilities can be denoted by $p_{\sigma ++},p_{\sigma +-}$ etc., with $p_{1++}=\frac12 p_1~,~p_{1+-}=\frac12 p_1~,~p_{1-+}=p_{1--}=0$ and similar for the other $\sigma $. On the level of the substates the observables $A^{(1)}$ and $A^{(2)}$ are classical observables, with $\bar A^{(1)}_\sigma =p_{\sigma ++}+p_{\sigma +-}-p_{\sigma -+}-p_{\sigma --},$ $\bar A^{(2)}_\sigma =p_{\sigma ++}-p_{\sigma +-}+p_{\sigma -+}-p_{\sigma --}$. (Since for half of the substates the probabilities vanish we could actually use a minimal set of eight substates.) When we will later discuss the time evolution of the ensemble, we will keep the relative probabilities for the substates, i.e. the ratios $p_{1++}/p_1~,~p_{2+-}/p_2$ etc.  fixed, according to the fixed entries in table \ref{table1}. This defines the notion of fixed probabilistic observables. Again, this discussion demonstrates how probabilistic observables can be implemented in standard classical statistics with classical observables. We emphasize, however, that such an implementation is not necessary and we will consider the probabilistic observables as genuine statistical objects. 

For probabilistic observables we encounter features well known from quantum mechanics. There are states where a given observable cannot have a sharp value. For example, the spin $A^{(2)}$ has a mean value zero in the states $(0)$ and $(\pi)$. For $p_1\equiv p_{(0)}=1~,~p_2=p_3=p_4=0$, we find a maximum variance for $A^{(2)}$, namely $\langle(A^{(2)})^2\rangle-\langle A^{(2)}\rangle^2=1$. On the other hand, for the state $(\pi/2)$, i.e. $p_1=p_2=p_4=0~,~p_3\equiv p_{(\pi/2)}=1$, the variance vanishes and $A^{(2)}$ has a sharp value. In analogy to quantum mechanics we will denote the states where some observable has zero variance, i.e. $\langle A\rangle^2=1$, as ``classical eigenstates'' for this observable. We call the value of the observable in such a classical eigenstate the ``classical eigenvalue''. For the spin $A^{(2)}$ we have two eigenstates, $(\pi/2)$ and $(-\pi/2)$, with respective eigenvalues $+1$ and $-1$. The setting of table 1 is analogous to two orthogonal spins in the quantum mechanics of a spin $1/2$ system. If one observable has a sharp value, the other has maximal uncertainty. 

We want to push the analogy with quantum mechanics even further and describe rotations in the plane spanned by the two spins within our setting of classical statistics. At this stage we encounter a problem. Rotating the pure state $(0)$ by an angle $\pi/4$, we should arrive at expectation values $\langle A^{(1)}\rangle =\langle A^{(2)}\rangle =1/\sqrt{2}$. This can not be realized in our system of four micro-states. Indeed, the sum of the components should be $\langle A^{(1)}\rangle+\langle A^{(2)}\rangle=\sqrt{2}$, while for an arbitrary probability distribution $\{p_\sigma \}$ we find the inequality
\begin{equation}\label{10}
\langle A^{(1)}\rangle+\langle A^{(2)}\rangle=p_1-p_2+p_3-p_4\leq 1.
\end{equation}
While the rotations of the state $(0)$ should lie on the circle in Fig. \ref{fig1}, the allowed macro-states of our four-state system are inside the square enclosed by the dashed lines. Only the four particular ``pure states'', where one of the $p_\sigma $ equals one, obey $\langle A^{(1)}\rangle^2+\langle A^{(2)}\rangle^2=1$. For $p_1=p_3=\frac12~,~p_2=p_4=0$ one has $\langle A^{(1)}\rangle=\langle A^{(2)}\rangle=1/2$ and therefore $\langle A^{(1)}\rangle^2+\langle A^{(2)}\rangle^2=1/2$. 

One may improve the situation by considering a classical system with eight microstates. For this purpose we add four more states denoted by $(\pi/4)~,~(-\pi/4)~,~(3\pi/4)~,~(-3\pi/4)$ - cf. the open circles in Fig.~\ref{fig1}. The mean values of the spins $A^{(1)}$ and $A^{(2)}$ in each of these four additional states are shown in the second part of Table \ref{table1}. The average values in an arbitrary macro-state are given by the eight probabilities $p_\sigma $ according to 
\ba\label{11}
\langle A^{(1)}\rangle&=&p_{(0)}-p_{(\pi)}\nonumber\\
&&+\frac{1}{\sqrt{2}}\left(p_{\left(\frac{\pi}{4}\right)}
+p_{\left(-\frac{\pi}{4}\right)}-p_{\left(\frac{3\pi}{4}\right)}
-p_{\left(-\frac{3\pi}{4}\right)}\right),\nonumber\\
\langle A^{(2)}\rangle&=&p_{\left(\frac{\pi}{2}\right)}-p_{\left(-\frac{\pi}{2}\right)}\\
&&+
\frac{1}{\sqrt{2}}
\left(p_{\left(\frac{\pi}{4}\right)}
-p_{\left(-\frac{\pi}{4}\right)}+p_{\left(\frac{3\pi}{4}\right)}
-p_{\left(-\frac{3\pi}{4}\right)}\right).\nonumber
\ea
A rotation by $\pi/4$ is now described by a change from the state $(0)$ to the state $(\pi/4)$. We define as ``classical pure states'' the ones which have one probability $p_{\bar\sigma }$ exactly equal to one and the others vanishing, $p_{\sigma \neq \bar \sigma }=0$. We have now eight pure states, and for all of them one observes $\langle A^{(1)}\rangle^2+\langle A^{(2)}\rangle^2=1$. The rotation of a pure state switches to another pure state. It is {\em not} realized by ``mixed states'' for which $\sum_{\sigma }p^2_\sigma <1$. (Note that  rotations for mixed states with $\langle A^{(1)}\rangle^2+\langle A^{(2)}\rangle^2\leq 1/2$ could, in principle, be realized by a suitable trajectory in the space of probability distributions $\{p_1,p_2,p_3,p_4\}.$)

For the system with eight states we can also define two further  two-level observables, namely $A^{(\pi/4)}$ and $A^{(-\pi/4)}$. These ``diagonal spins'' are specified by their mean values in the eight micro-states, as given by the entries in Table \ref{table2}. At this stage they are not obviously related to linear combinations of $A^{(1)}$ and $A^{(2)}$ - in fact, linear combinations are a priori not defined for the two-level observables that can only take values $+1$ and $-1$ for any measurement. We will only later introduce a concept of linear combinations such that the diagonal spins correspond to a rotation of the spins $A^{(1)}$ and $A^{(2)}$, similar to quantum mechanics. 

\begin{table}[h!tb]
\begin{tabular}{|c|cccc|cccc|}
\hline
~&$(0)$&$(\pi)$&$\left(\frac\pi 2\right)$&$\left(-\frac\pi 2\right)$&
$\left(\frac\pi 4\right)$&$\left(-\frac \pi 4\right)$&$\left(\frac{3\pi}{4}\right)$&$\left(-\frac{3\pi}{4}\right)$\\ \hline
$\bar A^{(\pi/4)}_\sigma $&
$\frac{1}{\sqrt{2}}$&$-\frac{1}{\sqrt{2}}$&$\frac{1}{\sqrt{2}}$&$-\frac{1}{\sqrt{2}}$&
$1$&$0$&$0$&$-1$\\
$\bar A^{(-\pi/4)}_\sigma $&$\frac{1}{\sqrt{2}}$&$-\frac{1}{\sqrt{2}}$&
$-\frac{1}{\sqrt{2}}$&$\frac{1}{\sqrt{2}}$&$0$&$1$&$-1$&$0$\\ \hline
\end{tabular}
\caption{Mean values for ``diagonal spins'' in different micro-states.}
\label{table2}
\end{table}

By suitable mixed states we can realize in the eight-state-system all expectation values of $A^{(1)}$ and $A^{(2)}$ within the dotted octogone in Fig. \ref{fig1}. Rotations for mixed states could now be achieved by appropriate $\{p_\sigma \}$ if $\langle A^{(1)}\rangle^2+\langle A^{(2)}\rangle^2\leq (2+\sqrt{2})/4$. For example, $p_{(0)}=\frac12~,~p_{\left(\frac\pi 4\right)}=\frac12$ yields $\langle A^{(1)}\rangle=\frac 12\left(1+\frac{1}{\sqrt{2}}\right)~,~\langle A^{(2)}\rangle=\frac{1}{2\sqrt{2}}$. For pure states, however, we still cannot realize the continuous rotations on the circle of Fig. \ref{fig1}, but only a discrete subset of rotations in units of $\pi/4$, corresponding to the $Z_8$-subgroup of the $SO(2)$ rotation group.

It is clear how to improve further by adding additional micro-states which interpolate closer and closer to the circle. The rotation problem for two spins can be solved by considering {\em infinitely many} different micro-states, each corresponding to a particular angle on the circle. With a finite number of $N$ micro-states we can come arbitrarily close to the continuous rotations by realizing a $Z_N$-subgroup. The full rotation group obtains in a well defined limit $N\to\infty$. 

The need of infinitely many micro-states for describing the rotation of a pure state in classical statistics should come as no surprise. By definition a pure state in classical statistics has zero variance for all {\em classical} observables. It is realized for probability distributions where one $p_\sigma $ equals one, while all others are zero. For classical pure states the statistical character of the probability distribution $\{p_\sigma \}$ is lost and each macro-state corresponds to one particular micro-state. The continuous rotation of a spin variable therefore requires a continuous family of micro-states. In other words, the continuous rotation of a planet is not described by different mixed states in a probability distribution, but just by different values of the angle which denotes the (deterministic) classical states which are pure states in a statistical sense. (Of course, pure states are only a very good approximation to the real statistical character of the planet.) The only thing that we have done in this section is a generalization of this situation from classical observables to the probabilistic two level observables. Nevertheless, this has a far reaching consequence: infinitely many classical states are needed for the description of a two-state quantum system. 

In summary of this section, a description of spin rotations requires a continuous manifold of microstates. In the simplest case this manifold is the circle $S^1$, parameterized by an angle $\varphi$. More complex manifolds may involve additional parameters of coordinates. In case of a manifold of microstates $S^1$, the probability of a particular state $\varphi$ or $(f_1,f_2)$ is given by $p_\sigma\equiv p(\varphi)=p(f_k)$. Here the cartesian coordinates'' $f_1,f_2$ obey $f^2_1+f^2_2=1$ and serve as a convenient alternative parameterization of the circle. For more complex manifolds $p(\varphi)$ corresponds to a marginalized probability, where one integrates over all other parameters or coordinates. ``Spins'' are described by a continuous family of probabilistic two-level observables $A(\psi)=A(e_k)~,~e^2_1+e^2_2=1$. Here $\psi$ or $(e_1,e_2)$ indicates the direction of the spin. The mean value of $A(e_k)$ in the state $f_k$ is given by 
\begin{equation}\label{15A}
\bar A_\sigma~\widehat{=}~\bar A_{f_k}(e_k)=\bar A_\varphi(\psi)=
\sum^2_{k=1}e_kf_k=\cos(\psi-\varphi),
\end{equation}
and the probability for finding for $A(e_k)$ the value $+1$ obeys
\ba\label{15B}
w_{\sigma +}~
&\widehat{=}&~w_{f_k,+}(e_k)=w_{\varphi,+}(\psi)\nonumber\\
&=&
\frac12(1+\sum_ke_kf_k)\nonumber\\
&=&\frac12\big(1+\cos(\psi-\varphi)\big).
\ea
The average value in the ensemble or macro-state reads
\ba\label{15B1}
\langle A\rangle~\widehat{=}~\langle A(e_k)\rangle&=&\sum_{\{f_k\}}p(f_k)\bar A_{f_k}(e_k)\nonumber\\
&=&\int^{2\pi}_0d\varphi p(\varphi)\cos (\psi-\varphi).
\ea

\section{Reduction of degrees of freedom}
In contrast to the infinitely many micro-states in classical statistics, it is impressive how quantum mechanics solves the rotation problem very economically: only two quantum states are needed, described by a two-component complex wave function. In fact, the classical solution to the rotation problem seems to be characterized by a huge redundancy. An infinite set of continuous probabilities $p_\sigma $ is employed to describe the expectation value of a spin observable, which can be characterized by only two continuous variables, the angle and the length $(\langle A^{(1)}\rangle^2+\langle A^{(2)}\rangle^2)^{1/2}$. One may be tempted to reduce the number of degrees of freedom by ``integrating out'' some of the micro-states, and assigning new ``effective probabilities'' $\tilde p_\sigma $ to the remaining micro-states. Such a ``coarse graining of states'' can indeed be done - but the price is that the effective $\tilde p_\sigma $ can also take negative values, or the sum of them may become larger than one. The effective probabilities can therefore no longer be considered as true probabilities.

We may demonstrate this by considering the system with eight classical states, $(0),(\pi),\left(\frac\pi2\right),\left(-\frac\pi2\right),\left(\frac\pi 4\right),\left(-\frac\pi 4\right),
\left(\frac{3\pi}{4}\right),\left(-\frac{3\pi}{4}\right)$, and ``integrate out'' the states $\left(\frac\pi 4\right),\left(-\frac\pi4\right),\left(\frac{3\pi}{4}\right),\left(-\frac{3\pi}{4}\right)$ in favor of new ``effective probabilities'' $\tilde p_{(0)},\tilde p_{(\pi)},\tilde p_{\left(\frac\pi 2\right)}$ and $\tilde p_{\left(-\frac\pi 2\right)}$ for the remaining four states. Let us first integrate out only the state $\left(\frac\pi 4\right)$. In order to keep the same expectation values for $A^{(1)}$ and $A^{(2)}$ according to eq. \eqref{11}, we need
\ba\label{12}
p'_{(0)}-p'_{(\pi)}&=&p_{(0)}-p_{(\pi)}+\frac{1}{\sqrt{2}}p_{\left(\frac\pi 4\right)},\nonumber\\
p'_{\left(\frac\pi 2\right)}-p'_{\left(-\frac\pi 2\right)}&=&p_{\left(\frac\pi 2\right)}
-p_{\left(-\frac\pi 2\right)}+\frac{1}{\sqrt{2}}p_{\left(\frac\pi 4\right)},
\ea
or
\ba\label{13}
p'_{(0)}&=&p_{(0)}+\frac{\alpha }{\sqrt{2}}p_{\left(\frac\pi 4\right)}~,~
p'_{(\pi)}+\frac{\alpha -1}{\sqrt{2}}p_{\left(\frac\pi 4\right)},\\
p'_{\left(\frac\pi 2\right)}&=&p_{\left(\frac\pi 2\right)}+\frac{\beta}{\sqrt{2}}
p_{\left(\frac\pi 4\right)}~,~p'_{\left(-\frac\pi 2\right)}=
p_{\left(-\frac\pi 2\right)}+
\frac{\beta-1}{\sqrt{2}}p_{\left(\frac\pi 4\right)}.\nonumber
\ea
For the sum this yields
\ba\label{14}
\sum&=&p'_{(0)}+p'_{(\pi)}+p'_{\left(\frac\pi2\right)}+p'_{\left(-\frac\pi 2\right)}\\
&=&p_{(0)}+p_{(\pi)}+p_{\left(\frac\pi 2\right)}+p_{\left(-\frac\pi 2\right)}+\sqrt{2}
(\alpha +\beta-1)p_{\left(\frac\pi 4\right)}.\nonumber
\ea
Consider now the pure microstate $p_{\left(\frac\pi 4\right)}=1~,~p_{(0)}=p_{(\pi)}=p_{\left(\frac\pi 2\right)}=p_{\left(-\frac\pi 2\right)}=0$. If we want $p'_{(\pi)}$ and $p'_{\left(-\frac\pi 2\right)}$ to be positive, we need $\alpha \geq 1~,~\beta\geq 1$. However, in this case we obtain $\sum \geq\sqrt{2}$. We can therefore not keep simultaneously $\sum\leq 1$ and all $p'_\sigma $ positive!

We may fix the coefficients $\alpha $ and $\beta$ as $\beta$ as $\alpha =\frac12~,~\beta=\frac12$. Integrating out also the states $\left(-\frac\pi 4\right)~,~\left(\frac{3\pi}{4}\right)$ and $\left(-\frac{3\pi}{4}\right)$ we can express the effective probabilities $\tilde p_\sigma $ for the four-state system in terms of the probabilities $p_\sigma $ of the eight-state system as 
\ba\label{15}
\tilde p_{(0)}&=&p_{(0)}+\frac{1}{2\sqrt{2}}
\left(p_{\left(\frac\pi 4\right)}+p_{\left(-\frac\pi 4\right)}
-p_{\left(\frac{3\pi}{4}\right)}-p_{\left(-\frac{3\pi}{4}\right)}\right),\nonumber\\
\tilde p_{(\pi)}&=&p_{(\pi)}-\frac{1}{2\sqrt{2}}
\left(p_{\left(\frac\pi 4\right)}+p_{\left(-\frac\pi 4\right)}-p_{\left(\frac{3\pi}{4}\right)}
-p_{\left(-\frac{3\pi}{4}\right)}\right),\nonumber\\
\tilde p_{\left(\frac\pi 2\right)}&=&p_{\left(\frac\pi 2\right)}+\frac{1}{2\sqrt{2}}
\left(p_{\left(\frac\pi 4\right)}-
p_{\left(-\frac\pi 4\right)}+p_{\left(\frac{3\pi}{4}\right)}-p_{\left(-\frac{3\pi}{4}\right)}
\right),\nonumber\\
\tilde p_{\left(-\frac\pi 2\right)}&=&p_{\left(-\frac\pi 2\right)}-\frac{1}{2\sqrt{2}}
\left(p_{\left(\frac\pi 4\right)}-p_{\left(-\frac\pi 4\right)}+p_{\left(\frac{3\pi}{4}\right)}
-p_{\left(-\frac{3\pi}{4}\right)}\right).\nonumber\\
\ea
It is easy to verify that also the observables $A^{(\pi/4)}$ and $A^{(-\pi/4)}$ keep the same expectation values after the reduction of degrees of freedom. We can now compute the expectation values of all four two-level observables by using eq. \eqref{1}, with $\bar A_\sigma $ given by the left half of Tables \ref{table1} and \ref{table2}, and $p_\sigma $ replaced by $\tilde p_\sigma $. The only memory that we have started from a system with eight microstates is the modified range for the effective probabilities $\tilde p_\sigma $. Since only the combinations $\tilde p_{(0)}-\tilde p_{(\pi)}$ and $\tilde p_{\left(\frac\pi2\right)}-\tilde p_{\left(-\frac\pi 2\right)}$ appear in the expectation values, all these statements hold actually for an arbitrary choice of $\alpha $ and $\beta$ in eq. \eqref{13}. The actual range for $\tilde p_\sigma $ depends on the choice of $\alpha ,\beta$ - for the choice \eqref{15} we have $\tilde p_\sigma \geq -1/(2\sqrt{2})$ and $\sum_{\sigma }\tilde p_\sigma \leq 1$. 

We may proceed one step further and also integrate out the states $(\pi)$ and $\left(-\frac\pi2\right)$. We denote the resulting effective probabilities by $\rho_1$ for the state $(0)$ and $\rho_2$ for the state $\left(\frac\pi 2\right)$. They are given by
\ba\label{16}
\rho_1&=&\tilde p_{(0)}-\tilde p_{(\pi)}=p_{(0)}-p_{(\pi)}\nonumber\\
&&+\frac{1}{\sqrt{2}}\left(p_{\left(\frac\pi4\right)}+p_{\left(-\frac\pi4\right)}
-p_{\left(\frac{3\pi}{4}\right)}-p_{\left(-\frac{3\pi}{4}\right)}
\right),\nonumber\\
\rho_2&=&\tilde p_{\left(\frac\pi 2\right)}-\tilde p_{\left(-\frac\pi 2\right)}=
p_{\left(\frac\pi2\right)}-p_{\left(-\frac\pi2\right)}\\
&&+\frac{1}{\sqrt{2}}\left(p_{\left(\frac\pi 4\right)}-p_{\left(-\frac\pi 4\right)}+p_{\left(\frac{3\pi}{4}\right)}
-p_{\left(-\frac{3\pi}{4}\right)}\right).\nonumber
\ea
In the reduced two-state-system the expectation values for the spins simply read
\ba\label{17}
\langle A^{(1)}\rangle&=&\rho_1~,~\langle A^{(2)}\rangle=\rho_2~,~\\
\langle A^{(\pi/4)}\rangle&=&\frac{1}{\sqrt{2}}(\rho_1+\rho_2)~,~\langle A^{(-\pi/4)}
\rangle =\frac{1}{\sqrt{2}}(\rho_1-\rho_2).\nonumber
\ea
The ambiguity associated to the choice of $\alpha $ and $\beta$ in the previous step has now disappeared and the $\rho_k$ are fixed uniquely in terms of the eight original $p_\sigma $. This uniqueness follows directly from eq. \eqref{17}, since the $\rho_k$ are associated to expectation values which do not change in the course of the reduction of degrees of freedom. One finds for the range of the effective probabilities $\rho_k$ for the reduced two classical states
\begin{equation}\label{18}
-1\leq\rho_k\leq 1~,~\sum_k\rho^2_k\leq 1,
\end{equation}
with $\sum_k\rho^2_k=1$ precisely for the eight original pure classical states. The two state system is the minimal system which can describe the two spins $A^{(k)}$. 

We could have started our reduction procedure with some other classical system with $2^M$ micro-states. Proceeding stepwise by reducing $M$ consecutively by one unit, one finally arrives again at the two state system, with effective probabilities $\rho_k$ obeying the constraints \eqref{18} and expectation values $\langle A^{(k)}\rangle=\rho_k$. Starting with infinitely many micro-states, $M\to\infty$, one finds that arbitrary values of $\rho_k$ obeying $\sum_k\rho^2_k\leq 1$ can be realized. This follows directly from the observation that the expectation values of $A^{(k)}$ can take arbitrary values within the unit circle, $\sum_k\langle A^{(k)}\rangle^2\leq 1$, and eq. \eqref{17}. Starting with a statistical system with finite $M$ leads to further restrictions on the allowed range of $\rho_k$. This range simply coincides with the allowed range for $\langle A^{(k)}\rangle$. For $M=3$ it is given by the dotted octogone in Fig. \ref{fig1}.

We may also investigate systems with a third independent two level observable $A^{(3)}$. The first step of our construction involves now six micro-states with mean values of the bi-modal observables shown in Table \ref{table3}.

\begin{table}[h!tb]
\begin{tabular}{|c|cc|cc|cc|}
\hline
~&$(+_1)$&$(-_1)$&$(+_2)$&$(-_2)$&$(+_3)$&$(-_3)$\\ \hline
$\bar A^{(1)}_\sigma $&$+1$&$-1$&$0$&$0$&$0$&$0$\\
$\bar A^{(2)}_\sigma $&$0$&$0$&$+1$&$-1$&$0$&$0$\\
$\bar A^{(3)}_\sigma $&$0$&$0$&$0$&$0$&$+1$&$-1$\\ \hline
\end{tabular}
\caption{Microscopic mean values for three ``orthogonal'' two-level observables}
\label{table3}
\end{table}

The solution of the three-dimensional rotation problem by infinitely many micro-states proceeds in complete analogy to the discussion above, with an interpolation of the pure states towards the unit sphere $S^2$. Also the reduction to a smaller number of states by ``integrating out'' some of the micro-states is analogous to the two-dimensional case. The minimal system has three effective states, with effective probabilities $\rho_k,k=1,2,3$, obeying again the restrictions \eqref{18} and $\langle A^{(k)}\rangle=\rho_k$. 

In summary, the manifold of microstates must contain the submanifold $S^2$, parameterized by two angles $(\vartheta,\varphi)$ or a three dimensional vector of unit length $(f_1,f_2,f_3)~,~\sum_kf^2_k=1$. In case of the minimal manifold $S^2$ the classical statistical systems are given by probability distributions $p(f_k)$ or $p(\vartheta,\varphi)$. Again, for extended manifolds of states we consider marginalization. We observe that $S^2$ corresponds to the manifold of normalized pure states for two-state quantum mechanics, i.e. the projective Hilbert space. Probability distributions on the projective Hilbert space have been discussed by \cite{MIS}, \cite{POB}, \cite{MP}, \cite{BB}. The reduction of redundant degrees of freedom maps the probability distributions on $S^2$ onto three real numbers, $p(f_k)\to\rho_k$, according to
\begin{equation}\label{22A}
\rho_k=\sum_{\{f_k\}}p(f_k)f_k=\int d\Omega~ p(\vartheta,\varphi)f_k(\vartheta,\varphi).
\end{equation}
From $\sum_kf_k(\vartheta,\varphi)\leq 1$ and $\int d\Omega p(\varphi,\vartheta)=1$ one  derives the inequality
\begin{equation}\label{28X}
\sum_k\rho^2_k\leq 1.
\end{equation}

For a system with infinitely many micro-states we can also define infinitely many two-level observables. A given spin may be denoted by a three-vector $\vec A$, which can take an arbitrary direction in the cartesian system defined by the orthogonal ``directions'' corresponding to $\rho_1,\rho_2,\rho_3$. We may characterize the direction of the spin $\vec A$ by the cartesian coordinates of a three dimensional unit vector $e_k,\sum_k e^2_k=1$. A more accurate notation for $\vec A$ would be $A(e_k)$, since the measurements of $\vec A$ will not yield three real numbers, but only one with values $+1$ and $-1$. The mean value of $A(e_k)$ in the micro-state $f_k$ is given by
\begin{equation}\label{22B}
\bar A_{f_k}(e_k)=\sum^3_{k=1}e_kf_k,
\end{equation}
and, using eqs. \eqref{22A}, \eqref{3}, the expectation value of $\vec A$ obeys the intuitive simple rue (cf. eq. \eqref{17})
\begin{equation}\label{19}
\langle \vec A\rangle\equiv \langle A(e_k)\rangle=\sum^3_{k=1}e_k\rho_k.
\end{equation}

This key ingredient of our formalism will be addressed more formally later. At this place we observe that the three numbers $\rho_k$ are given by the expectation values of three basis observables $A^{(k)}$
\begin{equation}\label{22C}
\rho_k=\langle A^{(k)}\rangle.
\end{equation}
We will see that these numbers specify the state of the quantum system completely. In other words, we may consider the quantum system as a subsystem of a larger classical system with infinitely many degrees of freedom. Its state can be characterized by the expectation values of a number of basis observables - the three spin observables $A^{(k)}$ in our case. The expectation values $\rho_k$ cannot all take arbitrary values between $-1$ and $+1$. In our case they are restricted by the condition $\sum_k\rho^2_k\leq 1$.

\section{Density matrix}
In the following we will concentrate on manifolds of micro-states for which the inequality \eqref{28X} holds. This is automatic for the minimal manifold $S^2$, but may restrict the probability distribution $p_\sigma$ for larger manifolds of micro-states. If eq. \eqref{28X} is obeyed, the expression \eqref{19} can be brought into a form familiar from quantum mechanics. We associate to each two-level observable $A(e_k)$ a $2\times 2$ matrix
\begin{equation}\label{20}
\hat A(e_k)=\sum_k e_k\tau_k,
\end{equation}
with $\tau_k$ the three Pauli matrices obeying the anticommutation relation $\{\tau_k,\tau_l\}=2\delta_{kl}$. Similarly, we may group the effective probabilities $\rho_k$ into a ``density matrix'' $\rho$,
\begin{equation}\label{21}
\rho=\frac12(1+\sum_k\rho_k\tau_k).
\end{equation}
In terms of these matrices the expectation values obey the quantum mechanical rule
\begin{equation}\label{22}
\langle A(e_k)\rangle=\text{tr} (\hat A(e_k)\rho).
\end{equation}

In a quantum mechanical language the ``operator'' $\hat A$ is precisely the (unit-)spin operator $\hat{\vec S}$ in the direction of $\vec e$. For the infinite system we may therefore switch to the familiar spin-notation $\vec A\to \vec S$. In conclusion, we have established that the classical system with infinitely many degrees of freedom precisely obeys all quantum mechanical relations for the expectation values as given by
\begin{equation}\label{23}
\langle\vec S\rangle=\text{tr}(\hat{\vec S}\rho).
\end{equation}
In particular, all relations following from the uncertainty principle are implemented directly. For a quantum mechanical two state system all information about the statistical state of the system is encoded in the density matrix, which obeys the usual relations tr $\rho=1~,~\rho_{11}\geq 0~,~\rho_{22}\geq 0~,$ tr $\rho^2\leq 1$, and tr $\rho^2=1$ for pure states. For our classical statistical system these relations follow from the definition \eqref{21} and the range for $\rho_k$ in eq. \eqref{18}. The spin operators $\hat{\vec S}$ are the most general hermitean operators for the quantum mechanical two-state system. On the level of expectation values of operators we have constructed a one to one mapping between the quantum mechanical two-state system and a classical system with infinitely many micro-states.

\section{Conditional correlation functions}
\label{conditional}
There are several ways to define correlation functions in a statistical system. The basic issue is the definition of a product between two observables $A$ and $B$. The product $AB$ should again be an observable. The (two-point-) correlation function is then the expectation value of this product, $\langle AB\rangle$. The choice of the product is, however, not unique. We have already demonstrated earlier how a quantum correlation function can arise from a classical statistical system \cite{3}.

\medskip\noindent
{\bf 1. Classical product}
For classical or deterministic observables the first candidate for a product between two observables $A,B$ is the pointwise or classical product
\begin{equation}\label{35A}
(A\cdot B)_\tau=A_\tau B_\tau
\end{equation}
In our approach, this definition is only possible on the level of substates $\tau$. For probabilistic observables on the level of micro-states, however, this type of product exists only for special cases, but not in general. We have argued in sect. II that the product of two probabilistic observables can be defined in a straightforward way only if they are ``comeasurable'', i.e. if the joint probabilities $w_{ab}$ for finding the value $\gamma_a$ for $A$ and $\gamma_b$ for $B$ are available, cf. eq. \eqref{9B}. In this case the product $AB$ in eq. \eqref{9B} is equivalent to the classical product \eqref{35A} on the substate level. In a quantum mechanical language, this type of structure can only be realized if the associated operators $\hat A$ and $\hat B$ commute \cite{CWN}, \cite{14A}.

In general, the property of comeasurability of two observables is lost when the substates are projected on the micro-states. Many different deterministic observables on the substate level, as characterized by their values $A_\tau$ in the substates $\tau$, are mapped into one and the same probabilistic observable. (The latter is characterized by $\bar A_\sigma$ for a two-level observable, and more generally by the spectrum $\gamma_a$ and the probabilities $w_a(\sigma)$.) Let us denote by $A_\tau$ and $A'_\tau$ two different classical observables on the substate level that are mapped into the same probabilistic observable $A$. While $\langle A\rangle=\langle A'\rangle$, the classical product with a different observable $B_\tau$ (which corresponds to a different probabilistic observable $B$) is, in general different for $A$ and $A'$
\begin{equation}\label{35B}
\langle A\cdot B\rangle\neq \langle A'\cdot B\rangle.
\end{equation}
In consequence, there cannot be a unique classical product between the probabilistic observables $A$ and $B$. We conclude that the classical product $\langle A\cdot B\rangle$ is a property of the system and its environment, since it involves information that is only available on the substate level. For measurements in the (sub-) system this information is not accessible. Predictions for measurements of the system have to be formulated on the level of micro-states involving an appropriate product for the probabilistic observables $A$ and $B$ which has to be determined. 

\medskip\noindent
{\bf 2. Pointwise product for probabilistic observables}

One possibility for a product of probabilistic observables is the ``probabilistic pointwise product'' that we denote with a cross, $A\times B$. It is defined by the multiplication of the mean values of $A$ and $B$ in every micro-state
\begin{equation}\label{24}
(\overline{A\times  B})_\sigma =\bar A_\sigma \bar B_\sigma .
\end{equation}
In other words, one multiplies the probability to find a value $x_A$ for $A$ with the probability to find $x_B$ for $B$,
\begin{equation}\label{25}
(\overline{A\times  B})_\sigma =\int dx_Adx_Bx_Ax_Bw^{(A)}_\sigma 
(x_A)w^{(B)}_\sigma (x_B).
\end{equation}
Thus $A\times B$ is again a probabilistic observable, with
\begin{equation}\label{26A}
w_\sigma(x)=\int dx_Adx_B\delta(x-x_Ax_B)w(x_A)w(x_B).
\end{equation}

Using the discrete formulation \eqref{8} for the two-level observables one has
\begin{equation}\label{26}
(\overline{A\times  B})_\sigma =w^{AB}_{+,\sigma }-w^{AB}_{-,\sigma },
\end{equation}
with $w^{AB}_{+,\sigma }$ the combined probability to find for the micro-state $\sigma $ a value $+1$ for $A$ and $+1$ for $B$, or $-1$ for $A$ and $-1$ for $B$, such that the sign of the product of values of $A$ and $B$ is positive. Similarly, $w^{AB}_{-,\sigma }$ obtains from the situations where the respective signs are opposite
\ba\label{27}
w^{AB}_{+,\sigma }=w^A_{+,\sigma }w^B_{+,\sigma }+w^A_{-,\sigma }w^B_{-,\sigma },\nonumber\\
w^{AB}_{-,\sigma }=w^A_{+,\sigma }w^B_{-,\sigma }+w^A_{-,\sigma }w^B_{+,\sigma }.
\ea
The probabilistic pointwise product of two two-level observables is again a two-level observable. 

The probabilistic pointwise product is commutative, and the corresponding pointwise correlation function equals the classical correlation function if $A$ and $B$ are classical observables. However, the pointwise product is not the product that leads to our definition of $A^2$ for the two-level observables, where $\langle A^2\rangle=1$ independently of $\langle A \rangle$. For the pointwise product one finds instead
\begin{equation}\label{28}
\langle A\times   A\rangle=\sum_\sigma  p_\sigma \bar A^2_\sigma  \leq 1.
\end{equation}
The saturation of the bound obtains only for the two pure classical states which correspond to the particular micro-states $\bar\sigma $ for which $\bar A=\pm 1$. This clearly indicates that the probabilistic pointwise product $A\times B$ cannot be used for the correlation between two measurements.

\medskip\noindent
{\bf 3. Conditional product}

The reason for this discrepancy is the implicit assumption that in eq. \eqref{27} the probabilities $w^A_\pm$ and $w^B_\pm$ are independent of each other. This does not reflect a situation where the product $AB$ describes consecutive measurements of first $B$ and subsequently $A$. Once the first measurement of $A$ has found a value $+1$, a subsequent measurement of the same variable should find a value $+1$ with probability one. Then $(\bar A^2)_\sigma =1$ follows independently of the value $\bar A_\sigma $ if the only allowed values are $\pm 1$, as for our spin variables. We therefore define a different product $A\circ  B$ which involves {\em conditional} probabilities. Eq. \eqref{27} is replaced by 
\ba\label{28a}
w^{AB}_{+,\sigma }&=&w^{A\circ B}_{+,\sigma}=
(w^A_+)^B_+w^B_{+,\sigma }+(w^A_-)^B_-w^B_{-,\sigma },\nonumber\\
w^{AB}_{-,\sigma }&=&w^{A\circ B}_{-,\sigma}=
(w^A_+)^B_-w^B_{-,\sigma }+(w^A_-)^B_+w^B_{+,\sigma }.
\ea
Here $(w^A_+)^B_+$ denotes the conditional probability to find a value $+1$ for the measurement of $A$ under the condition that a previous measurement of $B$ has yielded a value $+1$. With $(w^A_+)^A_+=1~,~(w^A_-)^A_-=1~,~(w^A_+)^A_-=0~,~(w^A_-)^A_+=0$ one now has $w^{AA}_{+,\sigma }=w^A_{+,\sigma }+w^A_{-,\sigma }=1~,~w^{AA}_{-,\sigma }=0$, such that $(\overline{A\circ  A})_\sigma =1$. More generally, we define the ``conditional product'' $A\circ B$ according to 
\begin{equation}\label{29A}
(\overline{A\circ B})_\sigma =w^{AB}_{+,\sigma }-w^{AB}_{-,\sigma },
\end{equation}
with the new definition \eqref{28a}. This product underlies our definition of $A^2,\langle A^2\rangle=\langle A\circ A\rangle =1$. The conditional product of two two-level observables is again a two-level observables, with probabilities given by eq. \eqref{28a}.

The definition of the conditional product $A\circ  B$ requires a specification of the conditional probabilities $(w^A_+)^B_+$ etc.. After the measurement of $B$ we know for sure that $B$ has the measured value, say $+1$. The probability of finding again $+1$ in a repetition of the measurement must be one. This is the property of a classical eigenstate of $B$, that we denote by $(+_B)$. We therefore take for the conditional probabilities
\ba\label{29}
(w^A_\pm)^B_+&=&\frac12(1\pm\langle A\rangle_{+B}),\nonumber\\
(w^A_\pm)^B_-&=&\frac12(1\pm \langle A\rangle_{-B}),
\ea
with $\langle A\rangle_{\pm B}$ the expectation value of $A$ in the pure classical states $(\pm_B)$, and $(-_B)$ the eigenstate of $B$ with eigenvalue $-1$. For our particular system corresponding to two-state quantum mechanics the conditional probabilities are actually independent of the micro-state $\sigma $. They depend only on properties of the states $(+_B)$ or $(-_B)$. We obtain
\ba\label{30}
(\overline{A\circ  B)}_\sigma &=&w^{AB}_{+,\sigma }-w^{AB}_{-,\sigma }
=\langle A\rangle_{+B}w^B_{+,\sigma }-\langle A\rangle_{-B}w^B_{-,\sigma }\nonumber\\
&=&\frac12(1+\bar B_\sigma )\langle A\rangle_{+B}-\frac12(1-\bar B_\sigma )\langle A\rangle_{-B}.
\ea
At this stage the conditional product is not obviously commutative. 

However, we note that whenever
\begin{equation}\label{36A}
\langle A\rangle_{-B}=-\langle A\rangle_{+B}
\end{equation}
holds, as in our case, one simply finds from eq. \eqref{30} 
\ba\label{36B}
(\overline {A\circ B})_\sigma  &=&\langle A\rangle_{+B}=\text{tr}(\hat A\rho_{B+})~,\nonumber\\
(\overline {B\circ A})_\sigma &=&
\langle B\rangle_{+A}=\text{tr}(\hat B\rho_{A+}),
\ea
independently of the micro-state $\sigma $. For our two state system the pure state density matrices are easily found
\begin{equation}\label{36C}
\rho_{B\pm}=\frac12(1\pm \hat B)~,~\rho_{A\pm}=\frac12(1\pm\hat A)
\end{equation}
such that 
\begin{equation}\label{36D}
\langle A\rangle_{+B}=\langle B\rangle_{+A}=\frac12\text{tr}(\hat A\hat B).
\end{equation}
This shows the commutativity of the conditional product of two probabilistic two-level observables
\begin{equation}\label{41Q}
(\overline{A\circ B})_\sigma=(\overline{B\circ A})_\sigma.
\end{equation}

It is instructive to compute the conditional product of two basis observables
\begin{equation}\label{41X}
A^{(k)}\circ A^{(l)}=
\left\{\begin{array}{lll}
1&\textup{for}&k=l\\
R&\textup{for}&k\neq l
\end{array}\right..
\end{equation}
Here $R$ is the ``random two-level observable'' which obeys for every micro-state $\sigma$
\begin{equation}\label{41Y}
\bar R_\sigma=0~,~w^R_{+,\sigma}=w^R_{-,\sigma}=\frac12~,~
(\overline{R^2})_\sigma=1,
\end{equation}
and therefore
\begin{equation}\label{41Z}
\langle R\rangle=0~,~\langle R^2\rangle=1.
\end{equation}
We stress that $R$ is different from the ``zero-observable'', which takes a fixed value $0$ in every micro-state. For an arbitrary two-level observable $A$ one finds the relation 
\begin{equation}\label{41ZA}
R\circ A=R.
\end{equation}
The random observable is a special two-level observable since no micro-state can be an eigenstate to $R$, the eigenvalues being $\pm 1$. The product $A\circ R$ is therefore not defined since it would require a projection on eigenstates of $R$ after a first ``measurement of $R$''. 

\medskip\noindent
{\bf 4. Conditional correlation}

Let us next compute the probability that a sequence of a first measurement of $B$ and a subsequent one of $A$ yields the results $(+,+)$,
\ba\label{41A}
W^{AB}_{++}&=&(w^A_+)^B_+w^B_{+,s}=\frac14(1+\langle A\rangle_{+B})
(1+\langle B\rangle_s)\nonumber\\
&=&\frac14(1+\frac14\textup{tr}\{\hat A,\hat B\})(1+\textup{tr}(\hat B\rho))
\ea
and compare it with the sequence in the opposite order
\begin{equation}\label{41B}
W^{BA}_{++}=(w^B_+)^A_+w^A_{+,s}=\frac14(1+\frac14\textup{tr}\{\hat A,\hat B\})
(1+\textup{tr}(\hat A\rho)).
\end{equation}
The probabilities for the sequences in different order are not equal. If we realize the probabilistic observables $A,B$ as classical observables on a substate level (with sharp values $A_\tau,B_\tau=\pm 1$ in any given substate $\tau$), we could also compute a ``classical probability'' $\tilde W^{AB}_{++}$ that both $A$ and $B$ ``have'' the value $+1$. This would be given by the sum of the probabilities for all states $\tau$ for which $A_\tau=B_\tau=1$. The order does not matter for the classical probability, $\tilde W^{AB}_{++}=\tilde W^{BA}_{++}$. Clearly, the probabilities $W^{AB}_{++}$ and $W^{BA}_{++}$ in eqs. \eqref{41A}, \eqref{41B} differ from $\tilde W^{AB}_{++}$. This demonstrates that our definition of the conditional probabilities $(w_+^A)^B_+$ etc. is not equivalent to the ``classical conditional probability'', which can be obtained from $\tilde W^{AB}_{\pm\pm}$ and appropriate marginalizations. While $W^{AB}_{++}\neq W^{BA}_{++}$ and $W^{AB}_{--}\neq W^{BA}_{--}$, the probability of finding for the product of the two measurements the value $+1$, namely
\begin{equation}\label{41C}
w^{AB}_{+,s}=W^{AB}_{++}+W^{AB}_{--}=\frac12(1+\frac14\textup{tr}\{\hat A,\hat B\}),
\end{equation}
does not longer depend on the order
\begin{equation}\label{41D}
w^{AB}_{+,s}=w^{BA}_{+,s}.
\end{equation}
This ``loss of memory of the order'' for the sum \eqref{41C} is the basis for the commutativity of $A\circ B$. 

The ``conditional correlation'' is defined as 
\ba\label{31}
\langle A\circ  B\rangle&=&\sum_\sigma  p_\sigma  (\overline{A\circ  B})_\sigma =
\langle A\rangle_{+B}w^B_{+,s}-\langle A\rangle_{-B}w^B_{-,s}\nonumber\\
&=&\frac12(1+\langle B\rangle)\langle A\rangle_{+B}-\frac12(1-\langle B\rangle)\langle A\rangle_{-B},
\ea
with $w^B_{\pm,s}=\sum_\sigma  p_\sigma  w^B_{\pm,\sigma }$ and $\langle B\rangle=\sum_\sigma  p_\sigma  \bar B_\sigma =\sum_\sigma  p_\sigma  (w^B_{+,\sigma }-w^B_{-,\sigma })=w^B_{+,s}-w^B_{-,s}$. For our orthogonal spin observables $A^{(k)}$ it has the simple property
\begin{equation}\label{33}
\langle A^{(k)} \circ A^{(l)}\rangle=\delta^{kl},
\end{equation}
since $\langle A^{(k)}\rangle_{\pm A^{(l)}}=0$ for $k\neq l$. The conditional correlation 
reflects the properties of two consecutive measurements. It may therefore be more appropriate for a description of real measurements than the probabilistic pointwise correlation.

A priori, the order of the measurements may matter, i.e. $\langle B\circ  A\rangle$ may differ from $\langle A\circ  B\rangle$, but based on eq. \eqref{41Q} we conclude that the (two-point) correlation is actually commutative
\begin{equation}\label{34}
\langle A\circ  B\rangle=\langle B\circ  A\rangle.
\end{equation}
We will  use in the next section the mapping to quantum mechanics and give a general expression of $\langle A\circ  B\rangle$ in terms of the anticommutator of the associated operators $\hat A,\hat B$, 
\begin{equation}\label{35}
\langle A\circ  B\rangle=\frac12 \text{tr}\big(\{\hat A,\hat B\}\rho)\big).
\end{equation}
From eq. \eqref{35} the commutativity of the conditional correlation is apparent. 

\vspace{2.5cm}\noindent
{\bf 5. Conditional three point function}

The commutativity of the conditional two point correlation does not extend to the conditional three point correlation. We first define the conditional product of three two-level observables as
\ba\label{36}
\overline{(A\circ  B\circ C)}_\sigma&=&
(w^A_+)^B_+(w^B_+)^C_+w^C_{+,\sigma}
-(w^A_+)^B_+(w^B_+)^C_-w^C_{-,\sigma}\nonumber\\
&-&(w^A_+)^B_-(w^B_-)^C_+w^C_{+,\sigma}
+(w^A_+)^B_-(w^B_-)^C_-w^C_{-,\sigma}\nonumber\\
&-&(w^A_-)^B_+(w^B_+)^C_+w^C_{+,\sigma}
+(w^A_-)^B_+(w^B_+)^C_-w^C_{-,\sigma}\nonumber\\
&+&(w^A_-)^B_-(w^B_-)^C_+w^C_{+,\sigma}
-(w^A_-)^B_-(w^B_-)^C_-w^C_{-,\sigma}.\nonumber\\
\ea
It is constructed in analogy to the conditional two point function and involves in an intuitive way the probabilities of finding for the measurements of $A,B,C$ the sequences $(+,+,+),(+,+,-),(+,-,+)\dots (-,-,-)$, weighted with the appropriate product of the measured values. After a measurement of $\pm 1$ of $C$ the observable $B$  is measured in the $(\pm_C)$ eigenstate of $C$, and after a second measurement of $\pm 1$ for $B$ the observable $A$ is measured in the $(\pm_B)$ eigenstate of $B$. For the orthogonal two-level observables eq. \eqref{36} yields 
\begin{equation}\label{42XA}
A^{(k)}\circ A^{(l)}\circ A^{(m)}=\delta^{kl}A^{(m)}+(1-\delta^{kl})R.
\end{equation}

The conditional three point function
\begin{equation}\label{42A}
\langle A\circ B\circ C\rangle=\sum_\sigma p_\sigma
(\overline{A\circ B\circ C})_\sigma
\end{equation}
obtains from eq. \eqref{36} by the replacement $w^C_{\pm,\sigma}\to w^C_{\pm,s}$. 
Similarly to eq. \eqref{31}, the conditional three point function can be expressed as a product of expectation values
\ba\label{37}
&&\hspace{-0.7cm}\langle A\circ  B\circ   C\rangle=\frac14
\Big\{\langle A\rangle_{+B}\\
&&\big[(1+\langle B\rangle_{+C})(1+\langle C\rangle)
-(1+\langle B\rangle_{-C})(1-\langle C\rangle)\big]\nonumber\\
&&+\langle A\rangle_{-B}\nonumber\\
&&\big[(1-\langle B\rangle_{-C})
(1-\langle C\rangle)
-\big[(1-\langle B\rangle_{+C})(1+\langle C\rangle)\big]\Big\}.\nonumber
\ea

The quantum mechanical computation in the next section shows that the conditional three point correlation can be expressed as 
\begin{equation}\label{38}
\langle A\circ  B\circ  C\rangle =\frac14\text{tr}\Big(\big\{\{\hat A,\hat B\},\hat C\big\}\rho\Big).
\end{equation}
It is therefore invariant under the exchange of $A$ and $B$, but not with respect to a change of the positions of $B$ and $C$ or $A$ and $C$. For the orthogonal spin observables one finds from eq. \eqref{38}
\begin{equation}\label{39}
\langle A^{(k)}\circ  A^{(l)}\circ  A^{(m)}\rangle=\delta^{kl}\langle A^{(m)}\rangle,
\end{equation}
in accordance with eq. \eqref{42XA}. We recall that all expectation values in eq. \eqref{37} are well defined in our setting with infinitely many micro-states, such that the computation of $\langle A\circ  B\circ  C\rangle$ can be done entirely within classical statistics. The non-commutativity is a consequence of the definition of the conditional product, which is adapted to a sequence of measurements in a given order. 

The conditional product is not associative. This can be shown most easily by using the products of two and three basis observables \eqref{41X}, \eqref{42XA}, which yield 
\ba\label{45A}
(A^{(k)}\circ A^{(l)})\circ A^m&=&
\left\{\begin{array}{lll}
A^{(m)}&\textup{for}&k=l\\
R&\textup{for}&k\neq l
\end{array}\right.\nonumber\\
&=&A^{(k)}\circ A^{(l)}\circ A^{(m)}.
\ea
On the other hand, $A^{(k)}\circ (A^{(l)}\circ A^{(m)})$ is not defined for $l\neq m$ since $A^{(l)}\circ A^{(m)}$ has no eigenstates, while for $l=m$ the result $A^{(k)}$ differs from eq. \eqref{45A}. The lack of commutativity of the product $A\circ B\circ C$ arises from the lack of associativity. Eq. \eqref{45A} can be generalized to arbitrary two-level observables
\begin{equation}\label{45C}
A\circ B\circ C=(A\circ B)\circ C. 
\end{equation}
The observable $A\circ B$ can be evaluated in eigenstates $\rho_{C+}$ and $\rho_{C-}$ of $C$ such that $(A\circ B)\circ C$ is well defined. The probability of finding $(A\circ B)\circ C=+1$ is given by
\begin{equation}\label{45D}
w^{(A\circ B)\circ C}_+=(w^{(A\circ B)}_+)^C_+w^C_{+,\sigma}+
(w^{(A\circ B)}_-)^C_- w^C_{-,\sigma}
\end{equation}
and similar for $w^{(A\circ B)\circ C}_-$. With
\ba\label{45E}
(w^{(A\circ B)}_+)^C_+&=&(w^A_+)^B_+(w^B_+)^C_++(w^A_-)^B_-(w^B_-)^C_+\nonumber\\
(w^{(A\circ B)}_-)^C_-&=&(w^A_+)^B_-(w^B_-)^C_-+(w^A_-)^B_+(w^B_+)^C_-
\ea
we indeed find for $w^{(A\circ B)\circ C}_+$ all terms in eq. \eqref{36} with a plus sign. Again, $A\circ (B\circ C)$ is in general not defined.

\medskip\noindent
{\bf 6. Measurements as operations}

For our two-state quantum system the density matrix for an eigenstate of the observable $A$ or $B$ is unique, given by eq. \eqref{36C}, and describing always a pure state. (This does not hold for more than two states.) We can describe the measurement process by a series of ``classical operations''. The first measurement of $C$ operates a mapping $\tilde C$
\begin{equation}\label{E8A}
\tilde C~:~\rho\to w^C_{+,s}\rho_{C+}-w^C_{-,s}\rho_{C-}=\tilde\rho_C,
\end{equation}
where $\tilde\rho_C$ is a weighted sum of density matrices, but not a density matrix itself. If this is the only measurement, the expectation value of $C$ obtains by taking a trace of $\tilde\rho_C$,
\begin{equation}\label{E8B}
\langle C\rangle=\text{tr}\tilde\rho_C=w^C_{+,s}-w^C_{-,s}.
\end{equation}
A second measurement of $B$ induces a mapping $\tilde B$
\ba\label{E8C}
\tilde B:&&\rho_{C+}\to (w^B_+)^C_+\rho_{B+}
-(w^B_-)^C_+\rho_{B-},\nonumber\\
&&\rho_{C-}\to (w^B_+)^C_-\rho_{B+}
-(w^B_-)^C_-\rho_{B-},
\ea
such that the sequence of two operations reads
\ba\label{E8D}
\tilde B\tilde C:~\rho\to\tilde\rho_{BC}&=&
\big[(w^B_+)^C_+w^C_{+,s}-(w^B_+)^C_-w^C_{-,s}\big]\rho_{B+}\nonumber\\
&+&\big[(w^B_-)^C_-w^C_{-,s}-(w^B_-)^C_+w^C_{+,s}\big]\rho_{B-}.\nonumber\\
\ea
Again, if the measurement chain is finished one takes the trace of $\tilde\rho_{BC}$ for the  evaluation of the expectation value of the products $B\circ C$, reproducing eq. \eqref{31}. The non-commutativity of the classical operations is manifest. For example, after the second step $\tilde\rho_{BC}$ is a linear combination of $\rho_{B+}$ and $\rho_{B,-}$, while $\tilde \rho_{CB}$ involves a linear combination of $\rho_{C+}$ and $\rho_{C-}$. This chain of operations can be continued. A third measurement of $A$ maps
\begin{equation}\label{49A}
\tilde A:\quad \rho_{B\pm}\to (w^A_+)^B_\pm\rho_{A+}-(w^A_-)^B_\pm \rho_{A-}
\end{equation}
such that taking a trace of $\tilde\rho_{ABC}$ after the third measurement reproduces eq. \eqref{42A} with $w^C_{\pm,\sigma}$ replaced by $w^C_{\pm,s}$, as appropriate for $\langle A\circ B\circ C\rangle$. 

A general physical measurement process, both for classical statistical systems and for quantum systems, involves three basic ingredients. (i) {\em Records} indicate the state of some measurement device (apparatus) after a measurement. In general, there may be $R_1$ possible values $m_1(r_1)~,~r_1=1\dots R_1$, for the record of the first device. In our case of two-level observables the record of an appropriate apparatus involves only one bit, $r_1=1,2~,~m_1(1)=+1~,~m_1(2)=-1$. 

(ii) {\em State reduction} describes the influence of the measurement on the state of the system. This is irrelevant if only one measurement is performed, but crucial for the outcome of a sequence of several measurements. Let us consider ``minimally destructive measurements''. For each given record $\bar m_1$ the state of the system becomes after the measurement an ``eigenstate'' with ``eigenvalue'' $\bar m_1$. This means that the measurement of $\bar m_1$ simply eliminates from  the ensemble all states which would have a non-zero probability that a repetition of the same measurement would yield a different  record $m_1\neq \bar m_1$. No further modification is made for the ensemble. The original state ``splits'' into $R_1$ different alternatives of ``histories'' which do not influence each other. A second measurement, with an apparatus with $R_2$ possibilities, yields new records $m_2(r_2)$. For a subsequent measurement this second step is performed for the $R_1$ alternative outcomes of the first measurement separately. A convenient way to visualize the situation is a sequence of two Stern-Gerlach type measurements, where the second measurement is performed by two identical devices placed in the two beams into which the incoming beam is split after the first measurement. State reduction after the combination of the two measurements produces $R_1R_2$ different alternatives. Arbitrary sequences can be constructed in this way. 

(iii) {\em Evaluation} of the value of the measured observable is some rule how the different records $m_1,m_2\dots$ are mapped to the value $V$ of the measured observable, which is a real number. In our case of a sequence of measurements of three two-level observables the value of the observable $A\circ B\circ C$ is given by $V=m_1m_2m_3$. For $m_j=\pm 1$ the value $V$ may take the value $\pm 1$. More generally, the spectrum of possible values for $V$ may consist of $R_t$ different possibilities, $r_t=1\dots R_t~,~V=m_t(r_t)$. Our prescription is general enough to include observables which are measured by a ``chain of individual measurements''. 

For a prediction of the probability $w_t$ for finding the value $m_t$ from the combination of a chain of several individual measurements the state reduction is a crucial ingredient. Indeed, $w_t$ obtains by summing the probabilities of all alternatives or histories for which the records $(m_1,m_2,\dots)$ are mapped into a given $m_t$ by the evaluation of the observable. The probability of a given history, labeled by $(m_1,m_2,\dots)$, multiplies the probability of finding $m_1$ (for a given state of the systems) with the probability of finding $m_2$ in the eigenstate with eigenvalue $m_1$ resulting from the state reduction of the first measurement and so forth. The prescription for the state reduction is, in general, not unique since many different states may be eigenstates with a given eigenvalue $m_1$. The definition of a ``combined observable'' as $A\circ B$ needs a specification of the appropriate state reduction. 

\medskip\noindent
{\bf 7. Choice of correlation function}

Several interpretations can be given for the use of conditional probabilities. One refers to the change of knowledge of the observer after the first measurement. The second one assumes that the physical state of the system has been changed after the first measurement through interaction with the apparatus. This does not involve an observer. Finally, one may take the point of  view that physical theories only describe probabilities for different possible histories, while reality has only one given history. If part of this given history is revealed, for example by the first experiment, the possibilities contradicting the outcome of the first experiment can be eliminated. In this view the essential part of a physical theory are the correlations between different events. These correlations do, in general, not involve an observer. From the mathematical point of view all those interpretations are described by the same conditional probability. In a quantum mechanical language this corresponds to the famous ``reduction of the wave function'' after the first measurement.

On the level of micro-states the classical correlation $\langle A\cdot B\rangle$ is not available since the joint probabilities are not defined. Both the probabilistic pointwise correlation $\langle A\times B\rangle$ and the conditional correlation $\langle A\circ B\rangle$ describe idealized measurements. The probabilistic pointwise correlation assumes that two measurements are completely uncorrelated on the level of the microstates. Suppose that the probability distribution for the micro-states singles out a particular micro-state for which the observable $A$ has no sharp value. Describing two consecutive measurements of $A$ by the probabilistic pointwise correlation corresponds to a situation where after the first measurement of $A$ the system relaxes such that the system has lost memory if the first measurement has found the value $+1$ or $-1$. In contrast, the conditional correlation keeps this memory. It idealizes that one has exactly an eigenstate of the measured observable after the first measurement. In a real measurement situation there will always be some uncertainty in the measured value and there are possible physical influences between the first and second measurement. This would result after the first measurement in a state that is not precisely an eigenstate. In principle, one could define modified correlation functions in order to account for such effects. Obviously, the process of performing a sequence of $n$ measurements and multiplying the measured numbers, and then averaging over many such sequences under identical conditions, has the necessary product properties for the definition of an $n$-point correlation. 

Our close association of the correlation functions with sequences of measurements underlines that the definition of the correlation function is not unique. In principle, a classical statistical system admits many different possible definitions of product structures and corresponding correlation functions. The most appropriate choice may actually depend on the detailed physical circumstances. Besides  the probabilistic pointwise product $A\times B$ and the conditional product $A\circ B$ we recall that the ``classical product'' $A\cdot B$ can be realized if the probabilistic observables are realized as classical observables on a substate level. On the level of substates $\tau$ the classical product \eqref{35A}, $(A\cdot B)_\tau=A_\tau B_\tau$, involves the sharp values $A_\tau,B_\tau$ of the observables $A$ and $B$ in the substate $\tau$. The classical product can be associated to the elimination of substates that have values of $B$ different from the value found in the first measurement. This state reduction needs, however, a specification of the precise observable $B_\tau$ that is measured, and not only of the associated probabilistic observable. Classical correlations therefore correspond to measurements where the properties of both system and environment are determined simultaneously. Such measurements do not correspond to measurements of the system properties, which only should employ information contained in the system, but no information about the precise state of the environment.

Furthermore, for the classical product Bell's inequalities can be directly applied and lead to contradiction with observation. This may be interpreted as experimental evidence that classical correlations should indeed not be used for a description of measurements of properties of an isolated (sub-) system. 
On the other hand, the conditional product yields precisely the prediction of quantum mechanics for the possible outcomes of two measurements. We will therefore postulate that two measurements should always be described by the correlation function $\langle A\circ B\rangle$ which we may call the ``quantum correlation''. This should also hold for situations where no clear time ordering of the measurements of the observables $A$ and $B$ is possible. 

Both the classical correlation $\kl A\cdot B\kr$ and the correlation $\kl A\circ B\kr$ are conditional correlations in the sense that they describe a way how possibilities contradicting the first measurement are eliminated. This demonstrates that the general notion of a conditional correlation is not unique. Any probabilistic theory must therefore not only specify a rule how expectation values of observables are calculated, but in addition also rules for the ``measurement correlation'' $\kl AB\kr_m$ which specify the outcome of measurements of pairs of observables. The various possibilities for definitions of conditional probabilities arise from the simple observation that it is not sufficient to state that all possibilities contradicting the first measurement are eliminated after the first measurement. This can be done in different ways. One also has to specify which information is retained and therefore available for the second measurement. The  classical correlation $\kl A\cdot B\kr$ can be used only if the precise observable $B_\tau$, which measures properties of the environment in addition to properties of the system, is specified for the first measurement. A good measurement, which does not destroy the isolation of the subsystem, should not require the knowledge of properties of the environment for a determination of the available information after the first measurement. It must be possible to specify the state of the subsystem after the first measurement by using only information which characterizes the properties of the subsystem. The correlation $\kl A\circ B\kr$ has precisely these properties. We therefore propose that for an optimal ``minimally destructive'' measurement in a quantum system the measurement correlation should be given as $\kl AB\kr_m=\kl A\circ B\kr$. 

\section{Quantum statistics from classical statistics}
So far we have shown important analogies between the quantum mechanics of a two state system and classical statistics with infinitely many micro-states. In this section we will argue that all aspects of quantum statistics can be described by the classical system. Quantum statistics appears therefore as a special setting within classical statistics, where a particular class of probabilistic observables is investigated and a particular correlation is used. Inversely, the formalism of quantum mechanics is a powerful tool for the computation of properties in classical statistical systems, as the conditional correlation functions. 

\bigskip\noindent
{\bf 1. Expectation values}

A first basic ingredient of quantum mechanics is a description of the rule for the computation of expectation values of observables. As before, we restrict the discussion to two quantum states. At any given time the information about the state of the system is encoded in the density matrix $\rho$, which is a hermitean $2\times 2$ matrix, $\rho=\rho^\dagger$ with $0\leq\rho_{11}\leq 1,0\leq\rho_{22}\leq 1$, tr $\rho=1$, tr $\rho^2\leq 1$. (Quantum mechanics provides also a law for the time evolution of $\rho$, to which we will turn in the next section.) Quantum mechanics makes probabilistic statements about the outcome of measurements. They are predicted by the expectation values for hermitean operators $\hat A$, according to $\langle A\rangle=$tr$(\hat A\rho)$. For the two state system, the only hermitean operators are the (unit) spin operators $\hat{\vec S}$ up to an overall multiplicative factor. (Pieces proportional to the unit operator may be added trivially.) We have already shown, eq. \eqref{23}, that the law $\langle\vec S\rangle=$ tr $(\hat{\vec S}\rho)$ is obeyed by the classical system with infinitely many micro-states. Also the density matrix with the required properties can be computed from the classical probability distribution $\{p_\sigma \}$, using the method of reduction of degrees of freedom. For the continuous family of classical spin observables we therefore have already established that the expectation values obey the quantum mechanical law. 

The quantum law for expectation value can be used whenever the expectation value of an observable can be written in the form
\begin{equation}\label{B10a}
\langle A\rangle=e_k\rho_k+e_0=\textup{tr}(\hat A\rho)~,~\hat A=e_k\tau_k+e_0.
\end{equation}
This holds for all probabilistic observables for which the probabilities to find a given value in the spectrum can be computed from $\rho_k$ by a linear relation. We will call such observables ``system observables''. ``Quantum observables'' have the additional property that $f(A)$ is represented by the operator $f(\hat A)$. This holds for the two-level observables associated to spins, but not for the random observable $R$. The quantum observables are a subclass of the system observables. 

\medskip\noindent
{\bf 2. Pure states}

A second basic concept in quantum statistics is the Hilbert space of states $|\psi\rangle$. They describe pure quantum states by complex two-component normalized vectors, $|\psi\rangle=\psi~,~\langle\psi |=\psi^\dagger$, with $\langle \psi|\psi\rangle=\psi^\dagger\psi=1$. Pure quantum states have a density matrix obeying tr $\rho^2=1$ or $\sum_k \rho^2_k=1$. (As we have seen, they correspond to the pure classical states where $\{p_\sigma \}$ has one value one and only zeros otherwise.) The overall phase of $|\psi\rangle$ is unobservable and therefore irrelevant. Only two real numbers are needed in order to describe the physical properties of $|\psi\rangle$, in correspondence to the two independent real numbers $\rho_k$ which remain under the condition $\sum_k\rho^2_k=1$. One can therefore construct a mapping between a pure density matrix $\rho$ and the associated state $|\psi\rangle$ up to an arbitrary phase $e^{i\varphi}$. The mapping is straightforward for diagonal $\rho$
\ba\label{41}
\rho=\left(\begin{array}{ll}
1&0\\0&0\end{array}\right)
&\leftrightarrow&
|\psi\rangle=e^{i\varphi}
\left(\begin{array}{l}
1\\0
\end{array}\right)~;~
\nonumber\\
\rho=\left(\begin{array}{ll}
0&0\\0&1
\end{array}\right)
&\leftrightarrow&
\psi=e^{i\varphi}
\left(\begin{array}{l}
0\\1
\end{array}\right).
\ea
Any hermitean $\rho$ can be diagonalized by a unitary $SU(2)$ transformation, $UU^\dagger=1~,~\rho_d$ diagonal,
\begin{equation}\label{42}
\rho=U\rho_d U^\dagger~,~|\psi\rangle=U|\psi_d\rangle.
\end{equation}
The second equation \eqref{42} defines the state $|\psi\rangle$ associated to $\rho$ with $|\psi_d\rangle$ associated to $\rho_d$ according to eq. \eqref{41}. 

This definition implies that expectation values of arbitrary operators can be computed in pure states as (we use $\langle\psi_d|=(1,0)$) 
\ba\label{43}
\langle A\rangle&=&\langle\psi|\hat A|\psi\rangle=\psi^\dagger\hat A\psi\nonumber\\
&=&\psi_d^\dagger U^\dagger\hat A U\psi_d=(U^\dagger\hat A U)_{11}\nonumber\\
&=&tr~(U\rho_d U^\dagger\hat A)=tr(\rho\hat A). 
\ea
We recover the standard quantum mechanics law for the computation of expectation values of observables in terms of ``probability amplitudes'' $\psi$. To every two component complex vector $\psi$ we can associate a pure state density matrix by a two step procedure: (i) normalize $\psi$ by a rescaling with a constant such that $\psi^\dagger\psi=1$, (ii) construct $\rho_{\alpha \beta}=\psi_\alpha \psi^*_\beta$. In particular, an associated density matrix exists for arbitrary linear combinations $\beta_1\psi_1+\beta_2\psi_2$. An explicit example for a classical probability distribution which corresponds to an entangled state within four-state quantum mechanics \cite{N} can be found in sect. XI.

In quantum mechanics the transition amplitude $M_{ab}$ between two pure states $|\psi_a \rangle$ and $|\psi_b\rangle$ is defined as $M_{ab}=\langle\psi_a |\psi_b\rangle$, and the transition probability obeys $w_{ab}=|M_{ab}|^2$. We will next establish that the transition probability is precisely the conditional probability discussed in the preceeding section
\be\label{44}
(w^A_\pm)^B_+=|\langle\pm_A|+_B\rangle|^2~,~(w^A_\pm)^B_-=
|\langle\pm_A|-_B\rangle|^2.
\end{equation}
Here the quantum states $|\pm_A\rangle$ are the eigenstates of the operator $\hat A$ with eigenvalues $\pm 1$, 
\be\label{45}
\hat A|+_A\rangle=|+_A\rangle~,~\hat A|-_A\rangle=-|-_A\rangle.
\end{equation}
In order to show eq. \eqref{44} we use the completeness of the Hilbert space which allows the insertion of a complete set of states
\ba\label{46}
1=\langle +_B|+_B\rangle&=&\langle +_B|+_A\rangle\langle +_A|+_B\rangle+
\langle+_B|-_A\rangle\langle-_A|+_B\rangle,\nonumber\\
\ea
and
\ba
&&\langle +_B|\hat A|+_B\rangle\nonumber\\
&&=\langle+_B|\hat A|+_A\rangle
\langle+_A|+_B\rangle +\langle +_B|\hat A|-_A\rangle\langle-_A|+_B\rangle\nonumber\\
&&=|\langle+_A|+_B\rangle|^2-|\langle-_A|+_B\rangle|^2. 
\ea
From the definition of the conditional probability and eq. \eqref{46} one obtains 
\ba\label{47}
(w^A_+)^B_+&=&\frac12(1+\langle A\rangle_{+B})=\frac12(1+\langle+_B|\hat A|+_B\rangle)\nonumber\\
&=&|\langle +_A|+_B\rangle|^2,
\ea
and similar for the other combinations in eq. \eqref{44}. 

\medskip\noindent
{\bf 3. Operator product}

A third crucial ingredient for quantum statistics is the definition of an operator product $\hat A\hat B$ and the determination of quantum correlations, as
\ba\label{40}
Re(\langle AB\rangle)&=&Re(\text{tr}\big(\hat A\hat B\rho)\big)=\frac12\text{tr}\big(
\{\hat A,\hat B\}\rho),\nonumber\\
Re(\langle ABC\rangle)&=&Re\big(\text{tr}(\hat A\hat B\hat C\rho)\big)=\frac12\text{tr}
\big((\hat A\hat B\hat C+\hat C\hat B\hat A)\rho\big)\nonumber\\
&=&\frac14\text{tr}\Big(\big(\big\{\{\hat A,\hat B\},\hat C\big\}+\big[[\hat A,\hat B],\hat C\big]\big)\Big)\nonumber\\
&=&\frac14\text{tr}\Big(\big(\big\{\{\hat A,\hat B\},\hat C\big\}+
\big\{\hat A,\{\hat B,\hat C\}\big\}\nonumber\\
&&-\big\{\hat B,\{\hat A,\hat C\}\big\}\big)\rho\Big). 
\ea
Since we have defined for the classical system the spin operators as $2\times 2$ matrices, we can, of course, define the matrix product and compute the quantum correlations \eqref{40} for $\hat A,\hat B, \hat C$ corresponding to arbitrary spin observables. Beyond this formal definition of the quantum correlations \eqref{40} we want to establish their close connection to the conditional correlations discussed in the preceeding section.

For this purpose we compute the expectation value of the anti-commutator of two operators in an arbitrary pure state $|s\rangle$. With
\ba\label{48}
\langle s|\hat A\hat B|s\rangle&=&\langle s|\hat A|+_B\rangle
\langle +_B|\hat B|s\rangle+\langle s|\hat A|-_B\rangle
\langle-_B|\hat B|s\rangle\nonumber\\
&=&|\langle +_B|s\rangle|^2\langle+_B|\hat A|+_B\rangle-
|\langle -_B|s\rangle|^2\langle -_B|\hat A|-_B\rangle\nonumber\\
&&+(\langle +_B|s\rangle\langle s|-_B\rangle
\langle -_B|\hat A|+_B\rangle -c.c.),
\ea
one finds the conditional correlation \eqref{31}
\ba\label{49}
\frac12\langle s|\{\hat A,\hat B\}|s\rangle&=&Re\big(\langle s|\hat A\hat B|s\rangle\big)\nonumber\\
&=&\langle A\rangle_{+B}w^B_{+,s}-\langle A\rangle_{-B}w^B_{-,s}\nonumber\\
&=&\langle A\circ B\rangle_s,
\ea
where $w^B_{\pm,s}=|\langle\pm_B|s\rangle|^2$. This establishes the relation \eqref{35} for any pure state density matrix. The extension to arbitrary $\rho$ uses the fact that $\rho$ can always be written as $\rho=w_1\rho_1+w_2\rho_2$ with $\rho_{1,2}$ pure state density matrices and real $w_{1,2}\geq 0,w_1+w_2=1$. With $|s_1\rangle,|s_2\rangle$ the pure states corresponding to $\rho_1,\rho_2$, one has
\ba\label{50}
&&\hspace{-1.5cm}\frac12\text{ tr }\big(\{\hat A,\hat B\}\rho\big)=\frac{w_1}{2}\langle s_1|\{\hat A,\hat B\}|s_1\rangle
+\frac{w_2}{2}\langle s_2|\{\hat A,\hat B\}|s_2\rangle\nonumber\\
&=&\langle A\rangle_{+B}(w_1|\langle+_B|s_1\rangle|^2+w_2|\langle+_B|s_2\rangle|^2)\nonumber\\
&-&\langle A\rangle_{-B}(w_1|\langle-_B|s_1\rangle|^2+w_2|\langle-_B|s_2\rangle|^2).
\ea
Using 
\be\label{51}
\frac12\big(1\pm \text{tr}
(\rho\hat B)\big)=w_1|\langle\pm_B|s_1\rangle|^2+w_2|\langle\pm_B|s_2\rangle|^2
\ee
this shows that the r.h.s. of eq. \eqref{50} coincides with the last eq. \eqref{31}. An analogous, but somewhat more lengthy computation establishes eq. \eqref{38} for the conditional three point correlation, and can also be used for higher correlation functions.

\medskip\noindent
{\bf 4. Quantum measurements}

A fourth corner stone of quantum mechanics is a rule how to express the possible outcome of measurements in terms of expectation values of observables. Such a rule is needed for every theory. For our classical statistical system we employ a rule based on conditional probabilities for consecutive measurements. It is the same as in quantum mechanics. We have shown how the conditional correlation functions in classical statistics can be expressed in terms of quantum correlation functions. Inversely, our computation provides a physical interpretation of quantum correlations in terms of the outcome of a sequence of measurements. Only the real part of the expectation values of products of operators can be measurable quantities. From eq. \eqref{40} we see how they can be related directly to conditional correlations. We note that the three point functions $Re(\langle ABC\rangle)$ does not simply correspond to one order of measurements (say first $C$, then $B$, last $A$), but rather to a linear combination of sequences in different orders, as given by the last equation in \eqref{40}. This is closely related to the term involving commutators in eq. \eqref{40}. The one to one correspondence between quantum correlations and conditional correlations closes the proof of equivalence between quantum statistics and classical statistics with infinitely  many micro-states. All measurable quantities can be computed in either approach.

The quantum pure states and the classical pure states are in one to one correspondence. Both can be parameterized by the coordinates on the sphere $S^2$, as given by the condition $\text{Tr}\rho^2=1$ or $\sum_k\rho^2_k=1$. From the point of view of the classical probability distribution $\{p_\sigma \}$ the classical pure states are sharp states with one $p_{\bar\sigma }$ equal to one - namely the one in the direction specified by the location of the state on $S^2$, and all other probabilities vanishing, $p_{\sigma \neq\bar \sigma }=0$. There is no statistical distribution on the level of $\{p_\sigma \}$. For the pure states the statistical character of the system arises therefore only from the notion of probabilistic observables. Only one spin has a sharp value in a given pure state, namely $\bar A_{\bar\sigma }=1$. It is the one which points in the direction corresponding to the location of the state on the sphere. All other spins have $(\bar A_{\bar\sigma })^2<1$ and therefore correspond to measurements with a statistical distribution of values $+1$ and $-1$. (For this counting the directions of the spin variables cover only half of the sphere, $S^{2}/Z_2$, and we have omitted the trivial extension $\bar A_{\bar\sigma }=-1$ for the spin opposite to the direction of the state.) In this setting the statistical character of quantum mechanics is genuinly  linked to the notion of probabilistic observables. We also note that the notions of classical eigenstates and classical eigenvalues are in direct correspondence to the quantum mechanical definition of eigenstates and eigenvalues. 

\medskip\noindent
{\bf 5. Operator algebra}

Finally, quantum mechanics has the useful structure of linear combinations of operators, $\lambda_1\hat A+\lambda_2\hat B$. It is compatible with the rule for expectation values
\be\label{52}
\langle\lambda_1A+\lambda_2B\rangle=\lambda_1\langle A\rangle+\lambda_2\langle B\rangle.
\ee
We can take this construction over to the two-level observables $A,B$ in the classical system with infinitely many degrees of freedom. For real $\lambda_{1,2}$ obeying $\lambda^2_1+\lambda^2_2=1$ the combination $\lambda_1A+\lambda_2B$ is again a two-level observable - the linear combination remains a map in the space of two-level observables. We can define rotated spins in this way. (Obviously, this is no longer possible for a finite number of micro-states, where the linear combination with arbitrary $\lambda_i,\lambda^2_1+\lambda^2_2=1$, is no longer defined. For finite $N$ the allowed $\lambda_i$ have to be restricted such that allowed spins are reached by the rotation - in our example with two spins the rotations have to be restricted to discrete $Z_N$-transformations.) We may relax the condition $\lambda^2_1+\lambda^2_2=1$ by defining formally the multiplication of an observable by a complex number $\lambda$ using the replacement $\bar A_\sigma \to\lambda\bar A_\sigma $, such that $\langle \lambda A\rangle=\lambda\langle A\rangle$. For real $\lambda>0$ this amounts to a change of units for the observables, replacing in eq. \eqref{5} $\delta(x\pm 1)\to\delta(x\pm\lambda)$. Multiplication with $-1$ corresponds to a map of the spin to the spin with opposite direction on the sphere. For real $\lambda$ all observables remain two-level observables with $\bar A^2_\sigma =\lambda^2$. The multiplication with $i$, or generally complex $\lambda$, remains formal and is not related to the outcome of possible measurements. It is, nevertheless, a useful computational tool since it gives to the space of observables the structure of a complex vector space. This is analogous to the multiplication of quantum states $|\psi\rangle$ by arbitrary complex numbers. It is needed in order to implement the vector-space structure of Hilbert space, even though physical states should be normalized, $\psi^\dagger\psi=1$. 

By a combination of rotations in the space of two-level observables with $\langle A^2\rangle=1$ and scalings we have defined arbitrary linear combinations  $A=\sum_ke_kA^{(k)}$ of the classical ``basis observables'' $A^{(k)}$. These quantum observables are represented by a complex three-component vector $\vec e=(e_1\dots e_3$). The expectation values are defined in the classical ensemble and obey $\langle A\rangle=\rho_ke_k~,~\langle A^2\rangle=e_ke_k$. This is in one to one correspondence with the operators $\hat A=e_k\tau_k~,~\hat A^2=e_ke_k$. The hermitean conjugation of a classical observable $A^\dagger$ is defined as $e_k\to e^*_k$. Measurements must yield real values such that measurable observables are $(A+A^\dagger)/2$. The most general operator in the Hilbert space of two-state quantum mechanics reads $\hat A=e_k\tau_k+e_0$. Using the unit observable every operator has its corresponding classical two-level observable $A=e_kA^{(k)}+e_0$. The possible outcomes of individual measurements of $A$ in the classical ensemble are given by the eigenvalues of the $2\times 2$-matrix $\hat A$.

The quantum mechanical operator product $\hat A\hat B$ can be mapped onto a ``quantum product'' of classical probabilistic quantum observables $AB$, as defined by the associated vector $\vec e$ and $e_0$. 
\be\label{63A}
e^{(AB)}_0=e^{(A)}_le^{(B)}_l~,~e^{(AB)}_k=i\epsilon_{klm}e^{(A)}_le^{(B)}_m.
\ee
With this product we can define an algebra of probabilistic quantum observables that is isomorphic to the algebra of quantum operators. On the level of the probabilistic observables one may at first sight wonder why one should introduce the particular product \eqref{63A}. However, we have seen already how to employ the quantum product for the computation of the outcome of a sequence of measurements in terms of conditional correlations. The expectation value of the quantum product $AB$ is closely related to the expectation value of the conditional product $A\circ B$ by eq. \eqref{35}. Another important use of the quantum product $AB$ is the discussion of the minimal value of the product of the dispersions for two observables. It can be expressed in terms of the commutator $AB-BA$ by the Heisenberg uncertainty relation. 

\medskip\noindent
{\bf 6. Beyond quantum observables}

Not all possible observables in a quantum system find a standard description in terms of quantum operators. Here an ``observable'' refers to a property of the system whose value (a real number) can be measured by some suitable apparatus. For an ``observable of the quantum system'' the spectrum of its possible measurement values should be determined by the properties of the quantum system, and the probabilities for finding a value within the spectrum should be computable in terms of the information characterizing the state of the quantum system (the density matrix $\rho$).

As an example for a measurement apparatus we consider a sequence of two Stern-Gerlach magnets, with directions of the magnetic fields rotated by 90 degree relative to each other. For an arbitrary polarized incoming beam the outcome will be four beams. For definiteness, they correspond to the spin values of $(S_z,S_x)=(+,+),(+,-),(-,+),(-,-)$, were $S_z$ refers to the spin after the first apparatus and $S_x$ to the spin after the second apparatus. We choose a two-level observable $R$ which take the value $V=1$ if $S_z$ and $S_x$ have the same sign, and $V=-1$ for opposite signs. The probability $w_+$ for finding $V=1$ can be measured by dividing the number of atoms in the two beams $(+,+)$ and $(-,-)$ by the number of atoms in the incoming beam, and similar for $w_-$. Thus the spectrum $V=\pm 1$ is known and the probabilities $w_\pm$ can be measured and computed for any incoming state - the observable $R$ is an observable of the quantum system. 

For our setting a quantum mechanical computation yields  $w_+=w_-=1/2$ for an arbitrary polarization of the outcoming beam. In other words, the observable $R$ has a vanishing expectation value for all states of the quantum system. The only quantum operator consistent with this property is $\hat A=0$. However, this operator has a spectrum with only one possible eigenvalue, namely zero, and not $(+1,-1)$ as appropriate for our two-level observable. If we request that the spectrum of possible measurement values of an observable should correspond to the spectrum of eigenvalues of an associated operator, we must conclude that not all observables of a quantum system can be described by quantum operators. 

Within our setting of probabilistic observables the observable $R$ finds a simple place. It is given by
\be\label{49X}
R=A^{(1)}\circ A^{(3)}
\ee
and equals the random observable discussed in the preceeding section. In a certain sense the ensemble of probabilistic observables is more complete than the ensemble of quantum operators, since arbitrary observables of the quantum system can be described. In contrast, the standard association of operators and observables in quantum mechanics covers only the quantum observables, which are a subclass of the more general system observables. The random observable $R$ is a system observable but not a quantum observable. Of course, the concept of a larger class of probabilistic observables can be implemented in quantum mechanics just as well as in classical statistics.

\section{Simple example: cartesian spins}
\label{simpleexample}
The basic ingredients for the reduction of a classical ensemble to a subsystem with quantum behavior can be understood in a simple example with a finite number of classical substates $\tau$. Of course, as we have seen in sect. \ref{spinrotations}, rotation symmetry can no longer be realized in such a system. We will discuss here three cartesian spins $S_x,S_y,S_z$, while continuous rotations of these observables are not defined.

On the substate level we consider eight substates labeled by $\tau=1,\dots,8$ or $\tau=(+,+,+),$ $(+,+,-)$, $(+,-,+)$, $(+,-,-)$, $(-,+,+)$, $(-,+,-)$, $(-,-,+)$, $(-,-,-)$. The spin observables have fixed values in each substate
\ba\label{S1}
S_x&=&\left\{\begin{array}{rll}
1&\textup{for}&\tau=1,3,5,7\\
-1&\textup{for}&\tau=2,4,6,8,
\end{array}\right.\nn\\
S_y&=&\left\{\begin{array}{rll}
1&\textup{for}&\tau=1,2,5,6\\
-1&\textup{for}&\tau=3,4,7,8,
\end{array}\right.\nn\\
S_z&=&\left\{\begin{array}{rll}
1&\textup{for}&\tau=1,2,3,4\\
-1&\textup{for}&\tau=5,6,7,8.
\end{array}\right.
\ea
In a direct product basis we can represent $S_x=(1\otimes 1\otimes \tau_3)~,~S_y=(1\otimes\tau_3\otimes 1)~,~S_z=(\tau_3\otimes 1\otimes 1)$. The classical ensemble is specified by the probabilities $p_\tau\geq 0~,~\{p_\tau\}=(p_1,\dots,p_8)~,~\sum_\tau p_\tau=1$. All classical observables of this system can be constructed from $S_k$ and the classical products $S_k\cdot S_l~,~S_k\cdot S_l\cdot S_j$. Beyond the unit observable one has a total number of seven independent observables, comprising the three spins $S_k$ and four observables $E_1=S_2\cdot S_3~,~E_2=S_1\cdot S_3~,~E_3=S_1\cdot S_2~,~E_4=S_1\cdot S_2\cdot S_3$. By a measurement of all seven expectation values $\kl S_k\kr~,~\kl E_i\kr$ one can determine all probabilities $p_1,\dots,p_8$. The joint probabilities for arbitrary pairs of observables $A_y$ and $A_z$ are available (with $A_z=(S_k,E_i)$ and $A^2_z=1)$ and we deal with a complete statistical system. 

Let us now define a subsystem with two properties: (i) Only the three cartesian spins $S_k$ are system observables, while the observables $E_1,\dots, E_4$ are considered as ``environment observables''. Measuremts in the subsystem cannot determine the expectation values $\kl E_i\kr$ for the environment observables. In turn, the information contained in $\kl E_i\kr$ cannot be used for predictions of properties of the subsystem which do not involve the environment. Since the information about the joint probabilities for the spins $S_k$ is directly related to the expectation values $\kl E_i\kr$ they are not available in the subsystem. The subsystem is described by incomplete statistics. (ii) The purity of the system is bounded by one, $P=\sum_k\rho^2_k\leq 1~,~\rho_k=\kl S_k\kr$. This restricts the most general classical probability distribution, as given by seven independent real numbers $p_1,\dots,p_7$, to a subspace obeying the inequality $P\leq 1$, with 
\ba\label{S2}
P&=&3-4(3p_1+2p_2+2p_3+p_4+2p_5+p_6+p_7)\nn\\
&&+4(3p^2_1+2p^2_2+2p^2_3+p^2_4+2p^2_5+p^2_6+p^2_7)\nn\\
&&+8(2p_1p_2+2p_1p_3+p_1p_4+2p_1p_5+p_1p_6+p_1p_7\nn\\
&&+p_2p_3+p_2p_4+p_2p_5+p_2p_6+p_3p_4+p_3p_5\nn\\
&&+p_3p_7+p_5p_6+p_5p_7).
\ea
(For example, $p_1=1~,~p_{\tau\neq 1}=0$ is not allowed since this would lead to $P=3$.) The purity constraint $P\leq 1$ allows us to construct the density matrix $\rho=(1+\kl S_k\kr\tau_k)/2$ such that $\kl S_k\kr=\textup{tr}(\hat S_k\rho)$, with $\hat S_k=\tau_k$. The subsystem shows many properties of a quantum mechanical system for three cartesian spins, provided we use the appropriate conditional correlation for the prediction of the outcome of sequences of measurements. 

After a first measurement has found $S_z=1$ we should eliminate all states for which $\kl S_z\kr\neq 1$. This implies for the state after this measurement $p_5=p_6=p_7=p_8=0~,~p_1+p_2+p_3+p_4=1$. Obviously, this requirement is not enough to fix the state uniquely, since we still have three free real numbers $p_1,p_2,p_3$. Different types of measurements correspond to different $(p_1,p_2,p_3)$. A ``classical measurement'' would be able to employ information concerning both the subsystem and the environment. A ``perfect classical measurement'' could retain the relative weights of the probabilities $p_1,p_2,p_3,p_4$ as they have been before the measurement. The state after the measurement would then be characterized by the ``classical rule'', according to which the three numbers $p_j$ are given by $p_j=p'_j/(p'_1+p'_2+p'_3+p'_4)~,~j=1,2,3$. Here $p'_j$ denote the probabilities in the state before measurement. The corresponding measurement correlation equals the classical (or pointwise) correlation. 

This procedure is not possible, however, for a measurement that is compatible with the isolation of the subsystem. The information needed for the computation of $(p_1,p_2,p_3)$ after the measurement according to the classical rule, namely $p'_j,j=1,\dots 4$, is more than what is contained in the three numbers $\rho_k=\kl S_k\kr$ which characterize the state of the subsystem. Furthermore, the probabilities for the new state, computed according to the classical rule, would not necessary obey the purity constraint $P\leq 1$ any longer. A simple counter example has for the original state before the measurement $p'_1=p'_5=p'_8=1/3$ with purity $P'=1/3$, where the classical rule would imply after the measurement $p_1=1,P=3$. 

A good measurement of pure substate properties should not involve environment information for the determination of the state of the subsystem after the measurement. In addition, the purity constraint $P\leq 1$ should be obeyed for the state after the first measurement. Therefore the state after the first measurement should again be characterized by three numbers $\rho_k$, with $P=\sum_k\rho^2_k\leq 1$. As we have seen already in sect. \ref{conditional}, these requirements fix uniquely the state after the first measurement. It must obey $\rho_1=\rho_2=0~,~\rho_3=1$, since any nonzero $\rho_1$ or $\rho_2$, together with $P\leq 1$, would imply $\rho_3<1$ and therefore $\kl S_z\kr< 1$, in contradiction with the measurement $S_z=1$. This ``quantum rule'' for the determination of $(p_1,p_2,p_3)$ after the first measurement implies $p_2=p_3=1/2-p_1$. The classical probabilities are not fixed completely since $p_1$ remains free within the interval $0\leq p_1\leq 1/2$. However, the undetermined part only concerns properties of the environment. We note the relations
\be\label{S3}
p_1+p_2=p_1+p_3=p_2+p_4=p_3+p_4=\frac12,
\ee
which imply that a subsequent measurement of $S_x$ or $S_y$ has equal probability to find values $+1$ or $-1$. We may have started before the measurement with a state where $p'_2+p'_4=0$ (as, for example, with the state $p'_1=p'_5=p'_8=1/3)$. Nevertheless, after the measurement one finds $p_2+p_4=1/2$. The perfect ``subsystem measurement'', which does not involve environment properties, leads to a change in the relative probabilities for the classical ensemble of subsystem plus environment. (In the example with $p'_1=p'_5=p'_8=1/3$ the ratio $(p'_2+p'_4)/p'_1=0$ changes to $(p_2+p_4)/p_1=1/2p_1)$. In other words, a good subsystem measurement necessarily leaves traces in the environment. Typically, these traces are not recorded, however.

Although the three cartesian spins serve as an instructive example for many of the crucial statistical properties of quantum systems, they do not reproduce all features of a two-state quantum system. Arbitrary linear combinations of quantum operators do not have corresponding classical observables in the ensemble with only eight states. For this purpose we have to extend the discussion to classical ensembles with infinitely many states, as we have discussed in sect. \ref{spinrotations}. 

\section{Time evolution}
\label{timeevolution}
In this section we discuss the time evolution in the classical statistical system. Assume that at some time $t_1$ the probability distribution is $\{p_\sigma \}$, and at some later time $t_2$ it has changed to a different distribution $\{p'_\sigma \}$. The observables are kept fixed and we want to study how their expectation values change. We may define ``transition probabilities'' $\tilde S_{\sigma \rho}$ such that (with summation over repeated indices)
\be\label{53}
p_\sigma (t_2)=\tilde S_{\sigma \rho}(t_2,t_1)p_\rho(t_1).
\ee
The transition matrix $\tilde S_{\sigma\rho}$ should conserve the unit sum, $\sum_\sigma  p_\sigma (t_2)=1$. By the process of reduction of degrees of freedom we can associate to $\{p_\sigma \}$ and $\{p'_\sigma \}$ effective probabilities for an effective three-state classical system, namely $\rho_k(t_1)$ and $\rho_k(t_2)$, $k=1\dots 3$. The transition matrix $\tilde S_{\sigma \rho}$ induces a reduced transition matrix $S_{kl}$ for the density matrix,
\be\label{54}
\rho_k(t_2)=S_{kl}(t_2,t_1)\rho_l(t_1).
\ee
It is related to $\tilde S_{\sigma\rho}$ by
\be\label{106A}
S_{kl}(t,t')=
\frac{\sum_{\sigma\tau\rho}\tilde S_{\sigma\tau}(t,t')p_\tau(t')p_\rho(t')\bar A^{(k)}_\sigma\bar A^{(l)}_\rho}{\sum_m\rho^2_m(t')}.
\ee

For the minimal manifold of micro-states $S^2,\sigma=(f_1,f_2,f_3),\sum_kf^2_k=1$, the condition 
\be\label{28x}
\sum_k\rho^2_k(t)\leq 1
\ee 
is preserved by the transformation \eqref{54} by construction. For more general manifolds of micro-states we assume that the condition \eqref{28x} holds for all times $t$. 

For the computation of expectation values for the spin observables and their conditional correlations at any given time $t$ one needs only to know $\rho(t)$. The reduced transition matrix $S_{kl}(t_2,t_1)$ is then sufficient for a description of the time evolution. We observe that many different transition matrices $\tilde S_{\sigma \rho}$ are mapped onto the same $S_{kl}$, such that actually only a limited amount of information about $\tilde S_{\sigma\rho}$ is needed. We can consider equivalence classes for $\tilde S_{\sigma\rho}$, where two transition matrices leading to the same $S_{kl}$ are considered to be equivalent. Similarly, equivalence classes for probability distributions $\{p_\sigma \}$ are characterized by $\rho_k$. 

Let us introduce the concept of purity, 
\be\label{66A}
P=\sum_k\rho^2_k. 
\ee
Then $P=1$ corresponds to pure states, $P<1$ to mixed states and $P=0$ to equipartition, where all $\rho_k=0$. (The equivalence class of equipartition contains the classical equipartition state $p_\sigma =1/N$.) The most general time evolution of the classical system can change the purity. We will concentrate first on the case where the purity is conserved. The unitary time evolution for two-state quantum mechanics will follow from this simple assumption.

\medskip\noindent
{\bf 1. Unitary quantum time evolution}

In fact, the conservation of the length of the vector $\rho_k$ implies that $S_{kl}$ is an orthogonal $O(3)$-matrix, $S_{kl}=\hat S_{kl}$, $\sum_l\hat S_{kl}\hat S_{ml}=\delta_{km}$. Arbitrary $O(3)$-transformations acting on the $\rho_k$ can be represented as unitary transformations acting on the density matrix $\rho$ (cf. eq. \eqref{21}) as
\be\label{55}
\rho(t_2)=U(t_2,t_1)\rho(t_1)U^\dagger(t_2,t_1).
\ee
This follows from the equivalence of $SU(2)$ and $SO(3)$ (up to a factor $Z_2$, with $\rho$ invariant under the $Z_2$ transformation). Parameterizing
\be\label{56}
U=e^{i\alpha _0}e^{i\alpha_m\tau_m}
\ee
one finds
\ba\label{57}
\hat S_{kl}&=&(1-2\sin^2\gamma)\delta_{kl}+2\sin^2\gamma~\beta_k\beta_l
\nonumber\\
&&+2\sin\gamma\cos\gamma~\epsilon_{klm}\beta_m,\nonumber\\
\gamma^2&=&\vec\alpha^2~,~\beta_m=\frac{\alpha_m}{\gamma}.
\ea

From eq. \eqref{55} the time evolution in two-state-quantum mechanics follows in a standard way. We may consider infinitesimal changes of time, for which we find the von-Neumann equation
\be\label{58}
\frac{\partial\rho}{\partial t}=-i[\hat H,\rho]~,~
\hat H=i\frac{\partial}{\partial t_2}U(t_2,t_1)
U^\dagger(t_2,t_1)=\hat H^\dagger.
\ee
Pure states obey then the Schr\"odinger equation with a hermitean Hamilton operator $\hat H$
\be\label{59}
i\frac{\partial}{\partial t}|\psi\rangle=\hat H|\psi\rangle.
\ee
With $\hat H=H_k\tau_k$ we can write eq. \eqref{58} as
\be\label{60}
\frac{\partial\rho_k}{\partial t}=2H_l\rho_m\epsilon_{lmk}
\ee
and compare with the general formula
\be\label{60A}
\frac{\partial\rho_k}{\partial t}=\frac{\partial S_{kl}}{\partial t}
S^{-1}_{lm}\rho_m.
\ee
For $S=\hat S$ we extract
\be\label{61}
H_k=-\frac14\frac{\partial\hat S_{jl}}{\partial t}\hat S^{-1}_{lm}\epsilon_{jmk},
\ee
which yields $\hat H$ in terms of $\hat S$. We observe that eq. \eqref{60} is consistent with eq. \eqref{60A} only for orthogonal matrices $S$ - otherwise the r.h.s. of eq. \eqref{60A} is not antisymmetric under the exchange of the indices $k$ and $m$ in the matrix multiplying $\rho_m$, as is the r.h.s. of eq. \eqref{60}.

The unitary transformation in quantum mechanics can easily be related to an appropriate time evolution of classical probabilities on the level of micro-states. Consider first the minimal manifold of micro-states $S^2$. It is sufficient that the classical time evolution of $p(f_s)$ is described by a rotation of the three-dimensional unit vector $(f_1,f_2,f_3)$ or a corresponding rotation of the angles $(\varphi,\vartheta)$. Let us consider a statistical system where the probability distribution at time $t_1,~p(\vartheta,\varphi;t_1)$, changes at some later time $t_2$ to 
\be\label{105A}
p(\vartheta,\varphi;t_2)=p(\vartheta',\varphi';t_1),
\ee
where $\vartheta'=\vartheta'(t_2,t_1,\vartheta,\varphi)$ is given by a time dependent rotation on $S^2$, and similar for $\varphi'$. We can then compute the time dependence of the elements of the density matrix
\ba\label{105B}
\rho_k(t')&=&\int d\Omega~p(\vartheta,\varphi;t_k)f_k(\vartheta,\varphi)\nonumber\\
&=&\int d\Omega~ p(\vartheta',\varphi';t_1)f_k(\vartheta,\varphi)\\
&=&\int d\Omega'~ p(\vartheta',\varphi',t_1)f_k\big(\vartheta(\vartheta',\varphi'),\varphi(\vartheta',\varphi')\big),\nonumber
\ea
where $\vartheta(\vartheta',\varphi',t_2,t_1)$ expresses the ``fixed angle'' $\vartheta$ in terms of the ``rotating angle'' $\vartheta'$. Since $f_k$ is a unit vector on $S^2$ one has
\be\label{105C}
f_k\big(\vartheta(\vartheta',\varphi'),\varphi(\vartheta',\varphi')\big)=
\hat S_{kl}(t_2,t_1)f_l(\vartheta',\varphi'),
\ee
with $\hat S_{kl}$ an orthogonal matrix depending on time. Insertion into eq. \eqref{105B} yields
\ba\label{105D}
\rho_k(t_2)&=&\hat S_{kl}(t_2,t_1)\int d\Omega'~p(\vartheta',\varphi';t_1)f_l(\vartheta',\varphi')\nonumber\\
&=&\hat S_{kl}(t_2,t_1)\rho_l(t_1)
\ea
As we have shown above, the orthogonal transformation \eqref{54} results in the unitary quantum evolution \eqref{58}, \eqref{59}. The generalization to extend manifolds of micro-states or to substates is straightforward. If the states of the extended manifold are characterized by $(\vartheta,\varphi)$ and additional parameters $\alpha$ (which are assumed to be invariant under rotations) the probability distribution $p(\vartheta,\varphi,\alpha,t)$ has to change according to eq. \eqref{105A}, with $\alpha$ kept fixed.

\medskip\noindent
{\bf 2. Decoherence and syncoherence}

We next consider the general case of the evolution \eqref{54} where $S_{kl}$ is not necessarily an orthogonal matrix. An arbitrary change of the vector $\rho_k$ can be written as a combination of an orthogonal transformation and a scaling, $S_{kl}=\hat S_{kl}d$. This adds to eq. \eqref{58} a scaling part
\be\label{62}
\frac{\partial \rho}{\partial t}=-i[\hat H,\rho]+D\left(\rho-\frac12\right)~,~
D=\frac{\partial\ln d}{\partial t}.
\ee
For negative $D$ the density matrix will approach equipartition, $\rho=\frac12~,~\rho_k=0$, as time increases. This describes {\em decoherence} of a quantum system. For positive $D$ the purity tends to increase
\be\label{63}
\partial_tP=2 D P.
\ee
For any arbitrary distribution $\{p_\sigma \}$ a classical pure state has the maximum possible purity, tr$\rho^2=1$. For positive $D$ the system has therefore a tendency to reach a pure state for large time.

In general, $S_{kl}$ may depend on $\rho_k$, and this also holds for $\hat H$ and $D$. The standard linear time evolution of quantum mechanics obtains only in the limit where $\hat H$ is independent of $\rho$ and $D$ vanishes. If $D$ depends on $\rho_k$, it will itself depend on time and we may write on effective evolution equation
\be\label{64}
\frac{\partial}{\partial t}D=\beta_D(\rho_k,D).
\ee
For tr$\rho^2=1$ a positive value of $D$ is forbidden by the general properties of the probability distribution. Indeed, for the minimal manifold of micro-states $S^2$ a pure state has the maximal possible value $\textup{tr}\rho^2=1$ - a pure state cannot get purer than pure. This follows directly from the definition of $\rho_k$, cf. eqs. \eqref{22A}, \eqref{28x} and generalizes to extended manifolds whenever it is possible to project them on $S^2$ such that eq. \eqref{28x} holds. We consider here only this type of systems. A positive value $D>0$ for $P=1$ would then imply a further increase of $P$, which is excluded. If in the vicinity of pure states $D$ is positive for ensembles with $P=\rho_k\rho_k<1$, we conclude that $\beta_D$ must have a zero for $D=0$ and $P=1$. If this fixed point is attractive for increasing $t$, a pure state will be approached asymptotically. Unitarity of the time evolution is then a simple consequence of the system approaching this fixed point for large $t$. We call this approach to a pure state {\em syncoherence}. A physical example is a mixed state of a an atom involving different energy levels. Due to radiative decay it can exchange energy with its environment and may finally end in the ground state. This is a pure state if the ground state is not degenerate, and pure states may also be reached for the degenerate case by appropriate ``experimental preparation''. 

The existence of fixed points for $P=0$ and $P=1$ is quite generic. The precise form of approach depends, of course, on the system. If $\beta_D$ admits a Taylor expansion for the fixed point at $P=1$ and $D=0$, the lowest order terms are
\be\label{65}
\beta_D=-aD+b(1-P),
\ee
where the coefficients may depend on $\rho_k/\sqrt{\text{Tr}\rho^2}$ and $H_k$. In the vicinity of the fixed point and for approximately constant $a$ and $b$ eq. \eqref{65} implies an exponential approach to the pure state,
\ba\label{66}
1-P&=&x_1e^{-\epsilon_1t}+x_2e^{-\epsilon_2t},\nonumber\\
D&=&\epsilon_1x_1e^{-\epsilon_1t}+\epsilon_2x_2e^{-\epsilon_2t},\\
\epsilon_{1,2}&=&\frac12(a\pm \sqrt{a^2-4b}),
\ea
provided $a>0~,~0<b<a^2/4$.

\medskip\noindent
{\bf 3. Hamiltonian quantum evolution}

If the fixed point with $D=0$ is approached for a sufficiently large time, we will encounter the standard linear unitary time evolution of quantum mechanics if $\hat H$ becomes independent of $\rho_k$ at the fixed point. Otherwise, the system would be attracted to a unitary, but non-linear extended version of quantum mechanics - a possibility that is highly interesting in its own right. We should note, however, that symmetries may enforce linear quantum mechanics. For example, if $SO(3)$ symmetry is realized at the fixed point, the Hamiltonian can depend on $\rho_k$ only via the invariant $\rho_k\rho_k$. This approaches a constant, and therefore the fixed point value of $\hat H$ has to be independent of $\rho_k$. 

The case where $\hat H$ becomes independent of $\rho_k$ at the fixed point for $P=1$ seems rather generic. It corresponds to the Hamiltonian evolution of quantum mechanical pure states \eqref{59}. For $\hat H$ independent of $\rho_k$ we may write
\be\label{11A}
\hat H=\sum_kH_k\tau_k+H_0,
\ee
where the coefficients $H_k,H_0$ do not depend on the quantum state or on $t$. Therefore $\hat H$ can be associated with an observable of the system \eqref{20}. Since $\hat H$ generates the time translation we infer from Noether's theorem that this observable is the conserved energy of the system. This demonstrates how the pure state fixed point is related to the isolation of the system from its environment in the sense that no energy is exchanged. 

In summary of this section, we find that the quantum mechanical time evolution can emerge naturally from a large class of time evolving probabilities $p_\sigma (t)$ \eqref{53}. The reduction to the time evolution of the density matrix \eqref{62} is always possible. Generic time evolutions may be attracted either to quantum mechanical equipartition, $\rho_k=0$, or to a pure quantum state. The asymptotic approach to the pure quantum state fixed point could provide an explanation why we can observe so many quantum systems in nature. Indeed, if we consider our system as a subsystem of a much larger system, the time evolution of the subsystem may allow for dissipation of energy into the larger system. Quite often, the lowest energy state is a pure state which may be approached for large time. A mixed state of atoms in various energy levels will after some time be found in the pure ground state if energy can be dissipated by radiation.

\section{Pseudo quantum systems}
\label{pseudoquantum}
We have seen how quantum mechanics arises from classical statistics in the limit of infinitely many micro-states, if probabilistic observables and conditional correlations are considered and the time evolution conserves purity. It is interesting to ask if ``approximate quantum behavior'' can be observed if the number $N$ of micro-states remains finite. The investigation of systems with finite $N$ may also be relevant for practical computations of quantum systems, in the sense that one may consider a series with increasing $N$ and take the limit $N\to\infty$ for which all quantities should converge to the quantities in the quantum system. We will call a classical statistical system with finite $N$ a ``pseudo quantum system'' if it fulfills the following criteria:
\begin{itemize}
\item [(i)] There are $N$ micro-states labeled by unit vectors $(f_k),k=1,2,3,~\sum_kf^2_k=1$, with probabilities $p_\sigma \equiv p(f_k)$. 
\item [(ii)] A group of discrete symmetry transformations $G_N$ acts in the space of $f_k$. It is a subgroup of $SO(3)$ and converges to $SO(3)$ in the limit $N\to\infty$. (For the concrete example in sect. 3 this subgroup is $Z_N$, but we consider here more general cases and three dimensional discrete rotations.)
\item [(iii)] One considers $N$ two-level probabilistic observables, labeled by unit vectors $(e_k),~\sum_ke^2_k=1$, i.e. $A(e_k)$. The symmetry group $G_N$ also acts on $e_k$, such that the scalar product $\sum_ke_kf_k$ is invariant. The mean values in the micro-states $\sigma =f_k$ are given by 
\be\label{A1}
\bar A(e_k)=\sum_kf_ke_k,
\ee
and the expectation values read
\be\label{A2}
\langle A(e_k)\rangle=\sum_{\{f_k\}}p(f_k)\sum_k(f_ke_k).
\ee
\end{itemize}
If we consider conditional correlations, these pseudo-quantum systems will converge to two-state quantum mechanics for $N\to\infty$. (Generalizations for quantum mechanics with more than two states are possible, but will not be considered in this section.)

We want to understand the differences between the pseudo quantum systems and quantum mechanics. For this purpose we first perform the reduction of the degrees of freedom to three effective micro-states, with effective probabilities $\rho_k$. This reduction should keep the expectation values of all observables $\langle A(e_k)\rangle$ unchanged. It can be achieved by
\be\label{A2a}
\rho_k=\sum_{\{f_k\}}p(f_k)f_k,
\ee
where the sum is over all micro-states. This guarantees that the expectation values of all spins can indeed be written as
\be\label{A3}
\langle A(e_k)\rangle=\sum_k\rho_ke_k,
\ee
verifying eq. \eqref{19}. We observe that the expression \eqref{A3} has no ambiguity and does not depend on which effective micro-state is selected while the others are integrated out. 

At this point the only difference to quantum mechanics is the restricted range of $f_k$, which results is a restricted range of $\rho_k$. This range has the geometry of a (three-dimensional) polygone with $N$ corners, where the corners are given by the vectors $f_k$ and correspond to the classical pure states. It approaches the sphere in the limit $N\to\infty$ as a result of $SO(3)$-symmetry. Thus the limiting $SO(3)$-symmetry guarantees that quantum mechanics is reached in the limit $N\to\infty$. The conditional correlations are defined for the pseudo quantum system just as for the quantum system. (The only difference may be a restricted number of observables $A(e_k)$.) The formalism of quantum mechanics can be applied to pseudo quantum systems, with the only restriction that the range of $\rho_k$ and therefore the number of pure states $|\psi\rangle$ is restricted - there are precisely $N$ different pure states $|\psi\rangle$ instead of a continuum. Also the number of observables may change from the continuous family of spins to a finite number $A(e_k)$, but this is not necessary.

These differences are necessarily reflected in the time evolution. For pure states, a unitary evolution is only possible for discrete steps $\tau_i$, corresponding to the allowed discrete symmetry transformations of the group $G_N$. Then the Hamilton operator becomes the transfer matrix. Alternatively, one may consider a continuous time evolution which does not respect the conservation of purity, such that $P=\sum_k\rho^2_k<1$ for times in the interval between the discrete time steps for which a pure state is transformed into another pure state, $\tau_i\leq t\leq \tau_{i+1}$. Unitarity is violated for these intermediate times, but restored whenever $t$ reaches $\tau_i$. It is therefore maintained in the average for long enough time in units of $\tau_{i+1}-\tau_i$. 

Pseudo quantum systems can only occur if the continuous symmetry $SO(3)$ is violated and reduced to a discrete subgroup $G_N$. Inversely, a classical statistical system with $SO(3)$-symmetry has necessarily infinitely many micro-states. Quantum mechanics arises whenever the time evolution of classical probabilities can be described by $SO(3)$-rotations, provided the appropriate two-level operators and conditional correlations are considered. In this sense it is not a very special situation within classical statistics. We emphasize that the $SO(3)$ rotations do not necessarily reflect the rotations in physical space, but may be more abstract isospin-type rotations. It is not necessary that the system is $SO(3)$-symmetric. Rather it is sufficient that the time evolution describes a {\em continuous} trajectory on $S^2$. For example, the trajectories may be $U(1)$-rotations, as for the quantum mechanics of a spin in a homogeneous magnetic field. Continuous rotations can also arise if the Hamiltonian has no continuous symmetry at all. 

\section{Realizations of probabilistic observables}
\label{realizations}
Probabilistic observables play an important role in the derivation of the laws of quantum mechanics from classical statistics presented in this paper. Two attitudes towards this concept are possible. One takes probabilistic observables as the basic concept. It may be motivated by the assumption that the description of reality is genuinly probabilistic. If the state of the world can only be described by probabilistic concepts, it seems natural that the basic notion of an observable should also be probabilistic. Taking this attitude, the probabilistic character of an observable is the genuine situation. Classical observables that take a sharp value in all micro-states of the system are then a special case, corresponding to an idealization. 

As an alternative, one may also follow an approach where classical observables are the basic objects. Probabilistic observables are then an effective concept that arises if several states are grouped together into a new intermediate state, which may then be treated as a micro-state on a higher level. This approach resembles the familiar concept of block spins. In this section we compare both concepts in our setting where quantum physics arises from classical statistics. We emphasize that our description of the two-state quantum system does not depend on how the probabilistic observables are implemented - either as ``fundamental'' or a ``composite'' objects.

\medskip\noindent
{\bf 1. Realization as classical observables}

We start with the implementation in terms of classical observables where the probabilistic observables appear as composite objects. In this case the micro-states of this paper correspond to the intermediate states. They are composed of substates, i.e. the ``true microscopic states'' for which the observables take fixed values. We have already briefly alluded to this concept in sects. II, III. Consider the spin observable $A^{(1)}$ or $A(e_k)=A(1,0,0)$. Since in every micro-state $(f_k)$ it is characterized by relative probabilities for values $\pm 1$, one needs a classical observable which can only take either the values $+1$ or $-1$ for any substate. One therefore needs at least two substates for any micro-state with $\bar A_{f_k}(1,0,0)\neq \pm 1$. The mean value in the micro-state $(f_k)$, $\bar A_{f_k}(1,0,0)$, is then given by the relative probabilities of the two substates. For a given $f_k$ these relative probabilities are fixed ``once and forever''. Our setting and the quantum mechanical time evolution do not describe situations where these relative probabilities between the substates change.

At this point we have derived the composite probabilistic observable $A(1,0,0)$ from a classical observable $A^{(C)}(1,0,0)$ which takes respectively the values $A^{(C)}_\tau=+1$ in one of the substates, and $A^{(C)}_\tau=-1$ in the other one. Denoting the relative probabilities of the two substates of the state $(f_k)$ with $w_+(f_k),w_-(f_k)=1-w_+(f_k)$, the probabilities of the substates are given by $p_+(f_k)=p(f_k)w_+(f_k)$ and $p_-(f_k)=p(f_k)w_-(f_k)$, and the mean value of $A(1,0,0)$ in the micro-state $(f_k)$ reads $\bar A_{f_k}(1,0,0)=w_+(f_k)-w_-{(f_k)}$. For the opposite spin, $A(-1,0,0)$, one finds $\bar A_{f_k}(-1,0,0)=w_-(f_k)-w_+(f_k)$. 

We next add a second two-level observable $A^{(2)}=A(0,1,0)$. Since the relative probabilities $w_\pm(f_k)$ are already fixed by the mean values $\bar A_{f_k}(1,0,0)$, we need a furher classical observable $A^{(C)'}$ that again takes values $+1$ or $-1$. Each substate needed for a description of $A(1,0,0)$ has to be divided again into two further substates, such that each state $(f_k)$ has now four substates. This process continues if we add the ``diagonal spins'' $A\left(\frac{1}{\sqrt{2}},\frac{1}{\sqrt{2}},0\right)$ etc.. For $N$ two-level-observables (counting $A(1,0,0)$ and $A(-1,0,0)$ separately) one needs $N/2$ classical observables $A^{(C)}_N$ and $2^{N/2}$ substates for every state $(f_k)$. 

More formally, the possible states of the ensemble can be characterized by $\big(f_k;\{\gamma(g_k)\}\big)$, where $f_k\in S^2~,~g_k\in S^2/Z_2$ (using $\gamma(-g_k)=-\gamma(g_k)$) and $\gamma(g_k)=\pm 1$ associates to every direction $g_k$ a separate discrete variable. The probabilities of these states read (in a discrete notation for finite $N$)
\be\label{83A}
p\big(f_k;\{\gamma(g_k)\}\big)=\prod_{\{g_k\}}
\big[\frac12\big(1+\gamma(g_k)f_kg_k\big)\big]p(f_k).
\ee
All observables $A(e_k)$ have a fixed value $+1$ or $-1$ in every state, given by $\gamma(e_k)$. (In other words, the observable $A(e_k)$ picks out a specific $\gamma(g_k=e_k)$ and is independent of all $\gamma(g_k\neq e_k)$.) Integrating out the substates yields
\be\label{83B}
\sum_{\{\gamma(_k)\}}
p\big(f_k;\{\gamma(g_k)\}\big)=p(f_k)
\ee
and
\ba\label{83C}
\bar A_{f_k}(e_k)&=&\sum_{\{\gamma(g_k)\}}
p\big(f_k;\{\gamma(g_k)\}\big)\gamma(e_k)\\
&=&\sum_{\gamma(e_k)=\pm 1}
\left[\frac12\big(1+\gamma(e_k)f_ke_k\right]\gamma(e_k)=f_ke_k,\nonumber
\ea
such that one recovers the micro-states $f_k$ and the probabilistic observables at an intermediate level. In principle, one could try to realize this situation by a ``hidden variable theory''. For $N\to\infty$ this would involve infinitely many discrete variables $\gamma(g_k)$ plus two continuous angular variables which take values on $S^2$ (i.e. $f_k$). Some law (deterministic or not) would have to reproduce the probability distribution \eqref{83A} for finding the values $\big(f_k;\gamma(g_k)\big)$ of the hidden variables.

While such a description of probabilistic observables in terms of classical observables is possible, it needs for large $N$ a very high number of subststates. In consequence, one encounters a very high degree of redundancy of the description by unobservable quantities. In addition, the fact that the relative probabilities for the substates do not change in the course of the time evolution may need some explanation. Part of the complexity arises in this case from our use of microstates with fixed distributions for the probabilistic observables. Omitting the microstates one can construct much simpler classical statistical ensembles that realize two-state quantum  mechanics. An explicit example for a classical statistical ensemble that describes two-state quantum mechanics together with its environment can be found in \cite{14A}.

\medskip\noindent
{\bf 2. Fundamental probabilistic observables}

Alternatively, we may consider the notion of probabilistic observables as fundamental. We may still formulate the probabilistic observables in terms of a ``basic observable'' which takes values $B=\pm 1$. However, one such observable will now be sufficient for a description of all $A(e_k)$. As a fundamental object, a probabilistic observable is {\em defined} by the relative probabilities $w_\pm (f_k)$ to observe $B=1$ or $B=-1$ in a given state $(f_k)$. Different probabilities $w_\pm(f_k)$ simply define different probabilistic observables. Instead of considering one fixed value of $w_\pm(f_k)$, a change of the relative probability for a given $(f_k)$ describes now the change from one observable to another.

The required relative probabilities $w_+(f_k)$ are easily computed for all two level observables $A(e_k)$ as 
\be\label{A4}
w_+(f_k;e_k)=\frac12\big(1+\bar A_{f_k}(e_k)\big)=\frac12(1+\sum_kf_ke_k). 
\ee
We may still introduce two substates $(f^+_k)$ and $f^-_k)$ for each micro-state $(f_k)$, and consider $B$ as a classical observable that takes the value $B=1$ for all substates $(f^+_k)$ and $B=-1$ for all substates $(f^-_k)$. However, the relative probability with which the substates are counted depends now on the observable $A(e_k)$ according to eq. \eqref{A4}. This dependence on $e_k$ appears manifestly in the expectation values
\ba\label{A5}
\langle A(e_k)\rangle&=&\sum_{\{f_k\}}\sum_{\gamma=\pm 1}\gamma\hat p_{e_k}(f_k,\gamma),\nonumber\\
\hat p_{e_k}(f_k,\gamma)&=&\frac12(1+\gamma\sum_kf_ke_k)p(f_k).
\ea
Here $\hat p_{e_k}(f_k,\gamma)$ is the effective probability with which the possible values of $B$, namely $\gamma=\pm 1$, are counted for every substate. It obeys $\hat p(f_k,\gamma)\geq 0$ and $\sum _{\{f_k\}}\sum_\gamma\hat p(f_k,\gamma)=1$. However, $\hat p$ depends now on the observable, i.e. on $e_k$. This is a major difference from the usual setting in classical statistics. It reflects the probabilistic nature of the observables, where part of the probability information is used for the definition of the observable - in our case the relative substate probability \eqref{A4}. We note in this context that eqs. \eqref{A4}, \eqref{A5} define positive semidefinite probabilities for $\sum_ke^2_k=1$. The scaling of observables is achieved in this formulation by a scaling of $B$, i.e. by a multiplication of the first eq. \eqref{A4} by $\lambda$. 

For fundamental probabilistic observables the correspondence between classical statistics entities and quantum mechanical objects becomes quite close. The basic variable $B$ in classical statistics can only take the values $\pm 1$, corresponding to the eigenvalues of the normalized spin operators in quantum mechanics and therefore to the possible outcome of a single measurement of the observables. The continuum of classical pure states on $S^2$ corresponds to the continuum of quantum mechanical pure states. The continuum of classical spin observables $A(e_k)$ corresponds to the continuum of normalized spin operators in two-state quantum mechanics. In the classical statistics setting the mixed states are described at this stage by infinitely many probabilities $p(f_k)$, while the density matrix $\rho$ in quantum mechanics needs only one probability $w$ to decompose $\rho=w\rho^{(1)}+(1-w)\rho^{(2)}$ into two pure state density matrices $\rho^{(1)}$ and $\rho^{(2)}$. In this respect, classical statistics remains redundant. It describes quantities $p(f_k)$ that cannot be determined by measurements of the two-level observables $A(e_k)$. The redundancy can be removed by integrating over the micro-states using eq. \eqref{A2a} 
\be\label{85A}
\hat p_{e_k}(\rho_k,\gamma)=\frac12(1+\gamma\sum_k\rho_ke_k).
\ee

The formula \eqref{85A} permits also a different interpretation. One may consider $B$ as the true observable of the system, with discrete values $+1$ or $-1$ in the two ``basic states''. The basic states are further characterized by ``external properties'', namely the ``state of the atom'' labeld by $\rho_k=\sqrt{P}f_k$, and the ``measurement orientation'', labeled by $e_k$. Thus a basic state can be parameterized by four angles, the purity and one discrete variable $(f_k,e_k,P;\gamma)$, with $f_k\in S^2,e_k\in S^2/Z_2,P\in [0,1],\gamma\in Z_2$. The probabilities for the two basic states obey
\be\label{86}
p(f_k,e_k,P;\gamma)=\frac12(1+\gamma\sqrt{P}\sum_kf_ke_k)=\frac12(1+\gamma\sum_k\rho_ke_k),
\ee
such that actually only the relative angle $\varphi$ between the atom-polarization and the apparatus orientation matters, i.e. $f_ke_k=\cos \varphi$. One has obviously
\ba\label{87}
&&\sum_{\gamma=\pm 1}p(f_k,e_k,P;\gamma)=1~,~
0\leq p(f_k,e_k,P;\gamma)\leq 1,\\
&&\langle B\rangle=\sum_{\gamma=\pm 1}\gamma p(f_k,e_k,P;\gamma)=\sqrt{P}\sum_k f_ke_k
=\sum_k\rho_ke_k.\nonumber
\ea

This point of view reflects precisely the setting of the Stern-Gerlach experiment, which splits an incoming polarized atom beam into two beams with different directions, corresponding to $B=\pm 1$. The probability of finding an atom in the $B=1$ direction only depends on the angle between  the polarization and the inhomogeneous magnetic field of the apparatus, as given by $\sum_kf_ke_k$, and on the degree of polarization, as given by $P$. It obtains from eq. \eqref{86} with $\gamma=1$. In this setting the time evolution of the ``atom state'' is described by the deterministic evolution equation for the ``external parameters'' $\rho_k$. The shift in the point of view as compared to the probabilistic observables $A(e_k)$ consists in attributing the information contained in $e_k$ to the basic state, rather than to the observable. The price for the simplicity of this picture is, of course, the explicit appearance of the ``measurement orientation'' in the relative probability for the $B=1$ and $B=-1$ states. The probability of finding $B=1$ or $B=-1$ not only depends on the state of the atom, but also on the state of the apparatus used for the measurements. 

In the formulation with substates, a given substate $\big(f_k,\{\gamma(e_k)\}\big)$ contains simultaneously the information about the values of {\em infinitely} many observables $A(e_k)$. The orientation of the measurement apparatus then decides which one of the $A(e_k)$ measured. In contrast, the picture with basis states has only {\em one} ``basis observable'' $B$. One has to specify the condition under which it can be measured through the measurement orientation $e_k$. No apparatus must actually be present - eq. \eqref{86} defines the outcome for all possible measurement directions. For an actual measurement, the orientation of the apparatus then decides which one of the $e_k$ applies for the given measurement situation. 

\section{Four state quantum system and entanglement}
\label{four}
Most of the conceptual issues how quantum mechanics emerges from a classical statistical setting can be described in the simplest system which corresponds to two state quantum mechanics. For keeping the discussion as simple as possible we have so far concentrated on this system. Certain  important features of quantum mechanics, as the phenomenon of entanglement, are visible, however, only in more complex systems, as four state quantum mechanics. This also applies to the inconsistencies of certain approaches to a classical implementation of quantum mechanics, which become apparent by applying the Kochen-Specker theorem. They only arise if three or more quantum states are involved. (No such inconsistency arises in our approach.) A classical statistical ensemble which corresponds to four state quantum mechanics and the phenomenon of entanglement have been discussed in ref. \cite{N}. For the sake of completeness of this paper we resume in this section certain key features of this work. 

\medskip\noindent
{\bf 1. Four-state quantum system}

Let us again consider a classical statistical ensemble, with microstates $\sigma$ labeled by a number of real parameters $f_k$ and probabilities $p_\sigma=p(f_k)\geq 0~,~\sum p(f_k)=1$. The manifold of micro-states parameterized by $f_k$ is now different from $S^2$ and will be specified below. We sometimes employ a discrete language corresponding to a finite number of values for $f_k$, such that a ``classical pure state'' has $p(f_k)=1$ for one particular microstate $f_k$, while $p$ vanishes for all other microstates. It is understood that we take a continuum limit with an infinite number of micro-states and a continuous vector $\vec f=(f_1,f_2\dots)$. Also the class of possible quantum observables will be extended beyond the two-level observables. Four state quantum mechanics has three linearly independent commuting operators and we will see that they correspond to the possibility to measure more than one bit simultaneously. 

We concentrate on possible measurements that can only resolve two bits. (In a quantum language this corresponds to two spins that can only have the values up or down.) For any individual measurement, the measurement-device or apparatus can only take the values $+1$ or $-1$ for bit $1$ and the same for bit $2$. In total there are four possible outcomes of an individual measurement, i.e. $(++),(+-),(-+)$ and $(--)$. We describe measurements of one bit again by two-level observables $A$ that are characterized by the probabilities $w^{(A)}_+(f_k)$ and $w^{(A)}_-(f_k)=1-w^{(A)}_+(f_k)$ to find a value $+1$ or $-1$ in any given microstate $f_k$. As before, the mean value of $A$ in a microstate $f_k$ reads
\be\label{qe1}
\bar A(f_k)=w^{(A)}_+(f_k)-w^{(A)}_-(f_k),
\ee
and the ensemble average obeys
\be\label{qe2}
\langle A\rangle=\sum\nolimits_{\{f_k\}}p(f_k)\bar A(f_k).
\ee

We concentrate first on three such ``two-level observables'', namely $T_1$ for the measurement of bit $1,~T_2$ for the measurement of bit $2$, and $T_3$ for the product of bit $1$ and bit $2$. Denoting by $W_{++},W_{+-},W_{-+}$ and $W_{--}$ the probabilities to measure in the ensemble the outcomes $(++),(+-),(-+)$ and $(--)$, one has 
\ba\label{qe3}
\langle T_1\rangle&=&W_{++}+W_{+-}-W_{-+}-W_{--}\nonumber\\
\langle T_2\rangle&=&W_{++}-W_{+-}+W_{-+}-W_{--}\nonumber\\
\langle T_3\rangle&=&W_{++}-W_{+-}-W_{-+}+W_{--},
\ea
such that $W_{++}$ etc. can be found from the average values of the three observables $\langle T_m\rangle$. For a classical eigenstate of the observable $T_1$ with eigenvalue $\langle T_1\rangle=1$ the probability for all states with $\bar T_1(f_k)<1$ must vanish. Such a pure state leads to $W_{-+}=W_{--}=0$.

Let us now specify our system. For the manifold of all microstates we choose the homogeneous space $SU(4)/SU(3)\times U(1)$. We parameterize the embedding space ${\mathbbm R}^{15}$ by the $15$ components $f_k$ of a vector $(k=1\dots 15)$. It is normalized according to $\sum_kf^2_k=3$, and obeys eight additional constraints that reduce the independent coordinates to six, as appropriate for the dimension of $SU(4)/SU(3)\times U(1)$. An easy way to obtain the constraints for $f_k$ employs a hermitean $4\times 4$ matrix $\tilde \rho$,
\be\label{qe5}
\tilde\rho=\frac14(1+f_kL_k)~,~f_k=\text{tr}(\tilde\rho L_k).
\ee
(Summation over repeated indices is always implied.) Here $L_k$ are fifteen $4\times 4$ matrices obeying
\ba\label{qe6}
L^2_k=1~,~\text{tr} L_k=0~,~\text{tr}(L_kL_l)=4\delta_{kl}.
\ea
They read explicitly (with $\tau_k$ the Pauli $2\times 2$ matrices)
\ba\label{qe5A}
L_1&=&\text{diag}(1,1,-1,-1)~,~L_2=\text{diag}(1,-1,1,-1)~,\nonumber\\
L_3&=&\text{diag}(1,-1,-1,1)~,~L_4=\left(\begin{array}{cc}\tau_1,&0\\0,&\tau_1
\end{array}\right),\\
L_5&=&\left(\begin{array}{cc}\tau_2,&0\\0,&\tau_2
\end{array}\right),
L_6=\left(\begin{array}{cc}\tau_1,&0\\0,&-\tau_1
\end{array}\right),
L_7=\left(\begin{array}{cc}\tau_2,&0\\0,&-\tau_2
\end{array}\right),\nonumber
\ea
with $L_8,L_9,L_{10},L_{11}$ obtained from $(L_4,L_5,L_6,L_7)$ by exchanging the second and third rows and columns, and $L_{12},L_{13},L_{14},L_{15}$ similarly by exchange of the second and fourth rows and columns. The matrix $\tilde \rho$ parameterizes the homogeneous space $SU(4)/SU(3)\times U(1)$ if it obeys
\be\label{qe7}
\tilde\rho=U\hat\rho_1U^\dagger~,~UU^\dagger=U^\dagger U=1~,~\hat\rho_1=\text{diag}(1,0,0,0),
\ee
for some appropriate unitary matrix $U$. This implies
\be\label{qe8}
\tilde\rho^2=\tilde\rho~,~\tilde\rho^2_{\alpha\alpha}\geq 0~,~\sum\nolimits_{\alpha}\tilde \rho_{\alpha\alpha}=\text{tr}\tilde\rho=1.
\ee

The observables $T_{1,2,3}$ are specified by $\bar T_m(f_k)=f_m~,~m=1,2,3$, which is equivalent to the specification of $w^{(T_m)}_\pm(f_k)$ in eq. \eqref{qe1}. Already at this stage we get a glance on the possibility of entanglement, since pure states with $f_1=f_2=0~,~f_3=-1$ will lead to $\langle T_1\rangle=\langle T_2\rangle=0$, $\langle T_3\rangle=-1$, and therefore to a correlation for opposite values of bit $1$ and bit $2,~W_{++}=W_{--}=0~,~W_{+-}=W_{-+}=\frac12$. We label these two-level observables by a real vector with components $e_k$, with $e_k(T_m)=\delta_{km},m=1\dots 3~,~k=1\dots 15$. The mean value of $T_m$ in a given microstate $f_k$ can then be written in the form (with $e_k\equiv e_k (T_m)$)
\be\label{qe9}
\bar T_m(f_k)=f_ke_k.
\ee
Similar to eq. \eqref{20} we represent an observable $A(e_k)$, labeled by $e_k$, in terms of a hermitean operator
\be\label{qe9A}
\hat A=e_kL_k~,~e_k(A)=\frac14\text{tr}(\hat AL_k),
\ee
with $\sum_ke^2_k=1$ for $\hat A^2=1$. In this language one finds $\bar T_m=\text{tr}(\hat T_m\tilde \rho)$, $\hat T_m=L_m$.  

We define a density matrix by
\be\label{qe11}
\rho=\frac14(1+\rho_kL_k)~,~\rho_k=\sum\nolimits_{\{f_k\}}p(f_k)f_k.
\ee
Eq. \eqref{qe2} yields the familiar quantum law for expectation values
\be\label{qe12}
\langle T_m\rangle=\text{tr}(\hat T_m\rho).
\ee
Much of the details of the classical probability distribution for mixed states cannot be resolved by measurements of $\langle T_m\rangle$ - only the entries of the density matrix $\rho_k$ matter. In contrast, for classical pure states only one microstate $f_k$ contributes, with $\rho_k=f_k~,~\rho=\tilde\rho$, and therefore $\rho^2=\rho$ as appropriate for a quantum pure state density matrix. 

The description of pure states in terms of wave functions $\psi_\alpha~,~\psi^\dagger\psi=1$, can be obtained from the density matrix in a standard way, $\tilde\rho_{\alpha\beta}=\psi_\alpha\psi^*_\beta~,~\psi_\alpha=U_{\alpha\beta}
(\psi_1)_\beta~,~(\psi_m)_\alpha=\delta_{m\alpha}$. This expresses the $f_k$ as a quadratic form in the complex four-vector $\psi_\alpha$,
\be\label{qe12B}
f_k=\psi^\dagger L_k\psi~,~\langle A\rangle=\psi^\dagger\hat A\psi,
\ee
and shows directly that only six components of $f_k$ are independent. The quantum mechanical wave function $\psi$ appears here as a convenient way to parameterize the manifold of microstates in classical statistics. The classical pure states are in one to one correspondence to the quantum pure states. For a pure state the purity $\rho_k\rho_k$ \eqref{66A} equals three.

\medskip\noindent
{\bf 2. Entanglement}

Let us next discuss a classical ensemble that realizes a typical entangled quantum state. We consider the wave functions
\be\label{qe14}
\psi_\pm=\frac{1}{\sqrt{2}}(\psi_2\pm\psi_3),
\ee
with associated pure state density matrices $\rho_\pm$. These states are eigenstates to $\hat T_3$ with eigenvalue $-1$. Writing
\be\label{qe16}
\rho_\pm=\tilde\rho_\pm=\frac14\big(1-L_3\pm(L_{12}-L_{14})\big),
\ee
we infer for the corresponding classical pure state $f_3=-1~,~f_{12}=\pm 1~,~f_{14}=\mp 1$, and all other $f_k$ vanishing. Thus $\langle T_1\rangle=\langle T_2\rangle=0$ implies that the values of bit $1$ and bit $2$ are randomly distributed, with equal probabilities to find $+1$ or $-1$. Nevertheless due to $\langle T_3\rangle=-1$, the product of both bits has a fixed value. Whenever bit $1$ is measured to be positive, one is certain that a measurement of bit $2$ yields a negative value, and vice versa. The two bits are maximally anticorrelated. We denote the {\em conditional probability} to find a value $\epsilon$ for bit $2$ if bit $1$ has been measured to have a value $\gamma$ by $p(\epsilon;\gamma)$. For our entangled state it obeys $p(1;1)=p(-1;-1)=0~,~p(1;-1)=p(-1;1)=1$. We will see below that for a typical entangled state further observables are strongly correlated or anticorrelated. 

Beyond $T_{1,2,3}$ we consider a set of fifteen basis observables $T_k$, $k=1\dots 15$. They are all two-level observables with spectrum $\pm 1$, specified by the mean value in a microstate $f_k$
\be\label{109}
\bar T_m(f_k)=f_m~,~m=1\dots 15.
\ee
The ensemble averages of the basis observables
\be\label{110}
\langle T_m\rangle=\sum_{\{f_k\}}p(f_k)\bar T_m(f_k)=
\sum_{\{f_k\}}p(f_k)f_m=\rho_m
\ee
characterize the quantum state, cf. eq. \eqref{qe11}. We can generalize eq. \eqref{qe9} for arbitrary $m$, with $e_k(T_m)=\delta_{km}$, and obtain $L_k$ as the quantum operators associated to $T_k$ by eq. \eqref{qe9A}.

Let us now describe the measurement of two spin observables with a relative rotation in the entangled state given by $\rho_-$ \eqref{qe16}. A rotated first spin observable $A(\vartheta)$ has the associated operator $\hat A(\vartheta)=\cos\vartheta L_1+\sin\vartheta L_8$, while a rotated second spin observable $B(\varphi)$ is associated to $\hat B(\varphi)=\cos\varphi L_2+\sin\varphi L_4$. This is most easily seen in a direct product basis where
\ba\label{111}
L_1&=&(\tau_3\otimes 1)~,~L_2=(1\otimes \tau_3)~,~L_3=(\tau_3\otimes\tau_3),\\
L_8&=&(\tau_1\otimes 1)~,~L_4=(1\otimes \tau_1)~,~L_{12}=(\tau_1\otimes\tau_1),\nonumber\\
L_6&=&(\tau_3\otimes\tau_1)~,~L_{10}=(\tau_1\otimes\tau_3)~,~L_{14}=-(\tau_2\otimes\tau_2).\nonumber
\ea
In this basis the entangled state density matrix $\rho_-$ \eqref{qe16} takes the intuitive form 
\be\label{112}
\rho_-=\frac14\big(1-(\tau_1\otimes\tau_1)-(\tau_2\otimes\tau_2)-(\tau_3\otimes\tau_3)\big).
\ee
All three spin components are maximally anticorrelated.

As we have discussed in sect. \ref{conditional} extensively, the product of two measurements of two spin observables is given by the conditional quantum correlation
\ba\label{113}
\langle BA\rangle_m&=&\big[(w^B_+)^A_+-(w^B_-)^A_+\big]w^A_{+,s}\nonumber\\
&-&\big[(w^B_+)^A_--(w^B_-)^A_-\big]w^A_{-,s}\nonumber\\
&=&\frac12\textup{tr}\big(\{\hat A,\hat B\}\rho\big).
\ea
Here the conditional probabilities are evaluated for minimally destructive measurements - details can be found in ref. \cite{CWN,14A}. For a general state one finds for the rotated spins the correlation
\ba\label{114}
\langle A(\vartheta)B(\varphi)\rangle_m&=&\cos \vartheta\cos\varphi\rho_3+\cos\vartheta\sin\varphi\rho_6\\
&+&\sin\vartheta\cos\varphi\rho_{10}+\sin\vartheta\sin\varphi\rho_{12}.
\nonumber
\ea
For the entangled state $\rho_-$ \eqref{qe16} one has $\rho_3=\rho_{12}=-1$, $\rho_6=\rho_{10}=0$ and therefore
\be\label{115}
\langle A(\vartheta)B(\varphi)\rangle_m=-\cos(\vartheta-\varphi)=\bar C_m(\vartheta-\varphi).
\ee
This is the same correlation as for quantum mechanics. 

In contrast, we may consider the ``classical correlation function''
\be\label{116}
\langle A(\vartheta)\cdot B(\varphi)\rangle =\bar C_{cl}(\vartheta-\varphi),
\ee
which could be defined if the probabilistic observables are realized as classical observables on the substate level, and if the conditional probabilities $W^{AB},W^{BA}$ in eqs. \eqref{41A}, \eqref{41B} are replaced by the ``classical probabilities'' $\tilde W^{AB}=\tilde W^{BA}$. In this case one can show Bell's inequality, which reads for our situation
\be\label{117}
|\bar C_{cl}(\vartheta_1)-\bar C_{cl}(\vartheta_2)|\leq 1+\bar C_{cl}(\vartheta_1-\vartheta_2).
\ee
The classical correlation \eqref{116} and the conditional correlation \eqref{115} are clearly different. Replacing $\bar C_{cl}$ in eq. \eqref{117} by $\bar C_m$ as given by eq. \eqref{115}, one finds that the inequality is violated for a range of angles, for example for $\vartheta_1=\pi/2$, $\vartheta_2=\pi/4$. This demonstrates again the crucial importance of the use of the appropriate correlation function for the description of the outcome of two measurements. 

\medskip\noindent
{\bf 3. Interference}

Other interesting quantum phenomena are the superposition of states and interference. Consider two pure quantum states that evolve in time according to
\be\label{qe17}
\psi_a=\frac{1}{\sqrt{2}}(\psi_1+\psi_2)e^{-i\omega_at}~,~
\psi_b=\frac{1}{\sqrt{2}}(\psi_1-\psi_2)e^{-i\omega_bt}.
\ee
The corresponding density matrices are time independent, $\rho_{a,b}=(1+L_1\pm L_4\pm L_6)/4$. Both states describe an eigenstate of the first bit, $\langle T_1\rangle =1$, whereas the second bit is randomly distributed, $\langle T_2\rangle=0$.  Due to the time dependent phase, the interference can be positive or negative for the superposition of both states, $\psi=\frac{1}{\sqrt{2}}(\psi_a+\psi_b)$. A quantum mechanical computation leads to a characteristic oscillation of $\langle T_2\rangle$,
\be\label{qe19}
\langle T_2\rangle=\psi^\dagger L_2\psi=\cos(\Delta t)~,~\Delta=\omega_a-\omega_b,
\ee
as known from the oscillation of a spin in the $z$-direction for a superposition of spin-eigenstates in the $x$-direction, with different energies for the positive and negative $S_x$ eigenvalues. A classical rotation $f_2=f_3=\cos(\Delta t)$ reproduces the ``interference pattern'' \eqref{qe19}. An evolution law $\partial_tf_2=\Delta f_5~,~\partial_tf_5=-\Delta f_2~,~f_3=f_2$, $f_7=f_5~,~f_1=1~,~f_k=0$ otherwise, has solutions leading to a density matrix which corresponds to the superposed state $\psi$,
\be\label{qe20}
\rho=\frac14\Big\{1+L_1+\cos(\Delta t)(L_2+L_3)-\sin(\Delta t)(L_5+L_7)\Big\}.
\ee
A classical statistical time evolution can yield the same dependence of expectation values as the quantum mechanical interference pattern. 

\medskip\noindent
{\bf 4. Fermions and bosons}

Our classical statistics setting can also describe identical bosons or fermions. We may identify the two bits with two particles that can have spin up or down,
\be\label{qe21}
\psi_1=|\uparrow\uparrow\rangle~,~\psi_2=|\uparrow\downarrow\rangle~,~\psi_3=|
\downarrow\uparrow\rangle~,~\psi_4=|\downarrow\downarrow\rangle.
\ee
If the particles are identical, no distinction between bit $1$ and bit $2$ should be possible. This requires that the system must be symmetric under the exchange of the two bits, imposing restrictions on the allowed probability distributions $p(f_k)$. The symmetry transformation corresponds to an exchange of the second and third rows and columns of $\tilde \rho$. On the level of the $f_k$ this amounts to a mapping $f_k\to f'_k~:~f_1\leftrightarrow f_2~,~f_4\leftrightarrow f_8~,~f_5\leftrightarrow f_9~,~f_6\leftrightarrow f_{10}~,~f_7\leftrightarrow f_{11}~,~f_{13}\leftrightarrow f_{15}$, while $f_3,f_{12}$ and $f_{14}$ remain invariant. Allowed probability distributions must obey $p(f'_k)=p(f_k)$. In particular, the allowed pure states are restricted by $f_1=f_2~,~f_4=f_8~,~f_5=f_9~,~f_6=f_{10}~,~f_7=f_{11}$, and $f_{13}=f_{15}$.

Consider the pure states $\psi_+$ and $\psi_-$ in eq. \eqref{qe14}. For both states the density matrix $\rho_\pm$  is compatible with the symmetry.  This does not hold for the density matrices corresponding to the states $\psi_2$ or $\psi_3$. In fact, linear superpositions of $\psi_+$ and $\psi_-$ are forbidden by the symmetry, a pure state $a\psi_-+b\psi_+$ must have $a=0$ or $b=0$. The symmetry requirement acts as a ``superselection rule'' for the allowed pure states or density matrices. For an arbitrary state vector $a\psi_-+b\psi_++c\psi_1+d\psi_4$ the symmetry of $\rho$ requires either $a=0$ or $b=c=d=0$. We observe that $\psi_-$ switches sign under the symmetry transformation as characteristic for a state consisting of two identical fermions. In contrast, the boson wave function $\psi=b\psi_++c\psi_1+d\psi_2$ is invariant under the ``particle exchange symmetry''.

\section{Probabilistic realism}
\label{probabilisticrealism}
We have explicitly constructed a classical statistical setting which realizes all laws of quantum mechanics. This construction is independent of the conceptual and philosophical interpretations of quantum mechanics. Nevertheless, it may have important conceptual consequences. In this section we argue that our setting is not in contradiction with physical realism, nor with locality in the sense that no signals traveling faster than light are needed. The quantum statistical systems are characterized, however, by a property of statistical ``incompleteness'', in the sense that joint probabilities cannot be used for the prediction of outcomes of measurements of arbitrary pairs of observables. This ``incompleteness'' is intrinsic for quantum systems - possible additional, more complete statistical information about joint probabilities is irrelevant for the outcome of measurements of the quantum observables. It can only specify some information about the ``environment'' of the quantum system. Statistical completeness for all observables, which means the availability and use of joint probabilities for measurement correlations of all pairs of observables, implies Bell's inequalities and therefore contradicts the observational evidence for quantum correlations. ``Local hidden variable theories'' usually assume statistical completeness. Such theories are not compatible with our setting.

From our point of view the most general description of physical reality is genuinely probabilistic \cite{GenStat}. Statements about reality concern expectation values and measurement correlations for observables. They are assumed to be, in principle, independent of the observer - the physical reality of correlations exists independently of an observer looking at them or not. In view of the presence of correlations in the cosmic microwave background emitted about 400 000 years after the big bang and concerning wavelengths of the size of the observable universe, it may indeed reasonably be assumed that such correlations are independent of a possible observation. This does not exclude that in some particular cases the correlations depend on the experimental setup - after all, the apparatus is part of the physical reality. 

We may quote the EPR-criterion \cite{EPR} for the existence of an element of physical reality: ``if, without in any way disturbing the system, we can predict with certainty (i.e. with probability equal to unity) the value of a physical quantity, then there exists an element of  physical reality corresponding to this physical quantity''. In the conceptual setting of ``probabilistic realism'' statistical correlations should be considered as possible ``elements of physical reality''. We will argue that in the typical EPR-case of two spatially separated spins, which are emitted from a spinless source and therefore have opposite directions, the physical reality concerns the maximal anticorrelation of the spins rather than the value of the spin of one of the particles. The essential statement of a physical theory describing the reality is then that the signs of the spins are opposite. This differs from the usual approach, where it is argued that at the moment when one of the spins is measured the value of the second spin is physical reality, and the second spin must therefore have this value even before the measurement if no signals from the measurement of the first spin can reach the second one. As is well known, this implies Bell's inequalities and leads to contradiction with quantum mechanics. In our view, the only element of physical reality which exists before the measurement is the maximal anticorrelation between the two spins. It is a priori not fixed if the necessary element of reality should refer to values of the spins or to correlations. In the present case it concerns the correlation. 

Of course, if one spin is measured, the value of the second one is fixed in consequence. After the measurement of one of the spins, one may eliminate all possibilities contradicting this measurement. This corresponds in quantum mechanics to the reduction of the wave function. We emphasize that this needs no exchange of signals and no fixed value of the second spin before the measurement. We could omit the reduction of the wave function, which is a pure tool of convenience, and only describe measurement correlations of events for the original wave function. A physical theory needs, of course, a specification how this correlation is calculated - in our approach as the conditional correlation in terms of the conditional probabilities.

Correlated systems cannot be separated into subsystems for which predictions can be made using only the information available in the subsystems. This is basic knowledge in any statistical system. In our approach it applies to the system of two spatially separated spins. Despite their separation, they cannot be treated as two independent systems of one particle with spin each. This would neglect the correlation. If one tries to do so nevertheless, one runs into conceptual contradictions. Some of the intuitive puzzles for the quantum mechanical system of two particles with total spin zero arise from the tendency to treat one of the particles as an isolated subsystem if it is separated sufficiently far from the other particle. However, due to the existence of correlations, the system always needs to be treated as a whole, even for arbitrarily large separation of the particles.

The possibility of nonlocal correlations is well known in statistical physics. This means that observables can be correlated even if they concern spatially separated regions and no signals can be exchanged between these regions. An example are macroscopic correlations between spins in a ferromagnet somewhat above the critical temperature, where the correlation length can reach a macroscopic size. Perhaps even simpler is the phenomenon of order. If the domains of magnetization are large enough, the measurement of the mean spin orientation in one region of the domain allows one to predict immediately the mean spin orientation in other regions of the domain. Therefore a correlation can be predicted even if measurements of the spin orientation are spatially separated in the sense that no signal can propagate between the different regions during the time of the measurement. 

Of course, the correlations must have been generated by local physical processes in the past. The original adjustment of the mean values of the spins into a given direction (within one of the ordered domains) must have proceeded by exchange process which can propagate at most with the speed of light. The analogue for the cosmic microwave background is the formation of correlations for metric fluctuations during the inflationary phase, which only later get separated to distances where signal exchange is no longer possible. Precisely the same happens for the spin correlations in the EPR-system with two particles with total spin zero. The correlation originates from the time of the decay of some spinless particle. Its persistence at later time is then simply a consequence of the conservation of angular momentum. 

What is then different between the correlations in quantum mechanics and the usual correlation in classical statistical systems, say in thermodynamics? The central issue concerns the question of completeness of the statistical system, rather than issues of locality or reality. We call a statistical system ``complete'' if joint probabilities are defined for all pairs of observables, and if the measurement correlations for all pairs of observables are predicted by the joint probabilities
\be\label{Z1}
\kl AB\kr_m=\sum_{a,b}a~b~p_{ab}.
\ee
Here $a$ and $b$ are the possible measurement values of the observables $A$ and $B$, respectively, and $p_{ab}$ denotes the joint probability that the measurement of $A$ yields $a$ and the measurement of $B$ yields $b$. We note that the property of completeness depends on the set of possible observables of the system.

It can be shown that eq. \eqref{Z1} implies Bell's inequalities \cite{CHSH}, \cite{B71}, \cite{CH}, \cite{CS}. One concludes that quantum statistical systems must be incomplete statistical systems. The measurement correlation is not given by eq. \eqref{Z1}, as we have already argued in sect. \ref{conditional}. In contrast, for a deterministic ``local hidden variable theory'' one assumes the existence of some set of hidden variables $\lambda$, which determine the values of the observables $A$ and $B$ as $a(\lambda)$ and $b(\lambda)$, respectively. Furthermore, it is assumed that for all $\lambda\in \Lambda$ a probability measure $d\rho$, with $\int_\Lambda d\rho=1$, is defined such that 
\be\label{Z2}
\kl AB\kr_m=\int_\Lambda d\rho a(\lambda)b(\lambda).
\ee
Eq. \eqref{Z2} implies Bell's inequalities. Local hidden variable theories are complete statistical systems. Indeed, we may denote by $\Lambda_{ab}$ the regions in parameter space for which the observable $A$ takes values in the internal $[a-\delta a, a+\delta a]$, and similarly $B$ is found in the internal $[b-\delta b b+\delta b]$, with $\delta a,\delta b$ specifying the ``resolution''. The joint probability is then given by
\be\label{Z3}
p_{ab}=\int_{\Lambda_{ab}}d\rho
\ee
and eq. \eqref{Z2} implies eq. \eqref{Z1} for $|\delta a|,|\delta b|\to 0$. (For a discrete spectrum of $A$ and $B$ one does not need $\delta a,\delta b,$ in this case $d\rho$ is directly given by $p_{ab}$.)

In our formalism with substates $\tau$ we could consider $\tau$ as hidden variables, since for every $\tau$ one has fixed values of the observables $A_\tau,B_\tau$ (corresponding to $a(\lambda),b(\lambda)$). The probability $p_\tau$ to find the substate $\tau$ specifies the joint probability. (In case of several states with the same $A_\tau,B_\tau$ one has to sum over all such states.) If the classical correlation
\be\label{Z4}
\kl A\cdot B\kr=\sum_\tau p_\tau A_\tau B_\tau
\ee
would define the measurement correlation, Bell's inequalities would follow. In sect. \ref{conditional} we have argued, however, that this correlation is not appropriate for statistical systems that describe isolated quantum systems since it measures properties of the environment of the system together with system properties.

On the level of microstates $\sigma$ the joint probabilities are not defined any longer. If the measurement correlation would be given by the probabilistic pointwise correlation
\be\label{Z5}
\kl A\times B\kr=\sum_\sigma p_\sigma\bar A_\sigma\bar B_\sigma,
\ee
this would again imply Bell's inequalities. Again, we have argued that the probabilistic pointwise correlation is not appropriate for the description of measurements of pairs of quantum observables. In contrast, the conditional correlation, which predicts the outcome of pairs of measurements, does not imply Bell's inequalities. Now the joint probabilities $p_{ab}$ are not used for the prediction of the outcome of measurements, since either they are not defined (on the level of microstates), or they do not describe system properties but rather also involve details of the environment which are not measured by a ``good quantum measurement'' (on the level of substates). We have seen that the conditional correlations precisely describe the correlations in quantum mechanics. The choice of the appropriate correlation function for the prediction of the outcome of measurements of pairs of observables is crucial for the understanding of quantum mechanics. 

\section{Conclusions}
\label{conclusions}
We have discussed classical statistical ensembles that exhibit all features of  two-state and four-state quantum systems. The quantum mechanical density matrix obtains by reduction of an infinity of classical micro-states to a few effective states. In turn, each micro-state can be obtained by a coarse graining of infinitely many substates. Most of the statistical information concerning the micro-states or substates is not needed for the description of the quantum system. It rather describes properties of the environment. All information relevant for the quantum system is retained by a ``coarse graining'' to a small number of effective states.

The minimal number of effective states depends on the observables which can describe an isolated (or approximately isolated) partial system as, for example, an atom in its environment. The expectation values of all observables of the partial system can be computed from the ``effective probabilities'' of the effective states which, in turn, are given by expectation values of suitable basis observables. We have constructed a density matrix from these expectation values. It has all the properties of the density matrix in quantum mechanics. In particular, the expectation values of observables of the partial system obey $\langle A\rangle=$tr$(\hat A\rho$), precisely the law of quantum mechanics. We have explicitly constructed the quantum mechanical operators $\hat A$ associated to classical spin observables $A$. They do not commute.

For pure states with tr$\rho^2=1$ one can ``take the root'' of the density matrix by introducing the quantum mechanical wave function $\psi$ in the usual way, with $\langle A\rangle=\psi^\dagger\hat A\psi$. The formalism of quantum mechanics, with probability amplitudes, superposition of states and interference is recovered. The quantum mechanical wave function appears here as a derived quantity rather than the fundamental object in quantum mechanics.

For two-state quantum mechanics the time evolution of the classical probability distribution is equivalent to the unitary time evolution of the density matrix only if the purity of the ensemble is conserved. This condition is generalized to quantum systems with more than two states in \cite{CWN,14A}. The unitary time evolution of the density matrix should be interpreted as a perfect isolation of the partial system described by the observables. As usual, a unitary evolution is described by a Hamilton operator $\hat H$. Since $\hat H$ is the generator of time translations it should correspond to the energy of the isolated partial system by virtue of the Noether theorem. A unitary time evolution of a pure state is described by the Schr\"odinger equation for $\psi$.

Our construction can be extended beyond the two-state and four-state quantum systems. For $M$ quantum states the manifold of micro-states parameterized by $f_k$ corresponds to the homogeneous space $SU(M)/SU(M-1)\times U(1)$, while the discussion of this paper was mainly restricted to $M=2$ where $SU(2)/U(1)$ parameterizes the sphere $S^2$. Furthermore, an explicit discussion of the phenomena of entanglement, superposition and interference within classical statistics is given for $M=4$ in sect. \ref{four}. For identical ``particles'' this also accounts for the difference between fermions and bosons. We observe that the restriction to a manifold of micro-states $SU(M)/SU(M-1)\times U(1)$ is not necessary. The latter is simply the minimal manifold needed in order to implement an unitary continuous time evolution. One may embed this manifold into a larger manifold of classical states. Then it appears as a projection of the larger ensemble on the minimal manifold of micro-states. The probability distribution on the minimal manifold of micro-states carries all the information needed for the expectation values of the observables of the ``isolated system'', plus irrelevant additional information if the state is mixed.

The unitary time evolution of quantum mechanics appears as a special case of a wider class of time evolutions of the classical ensemble. We argue that the special case of the unitary evolution of pure states corresponds to a partial fixed point of the more general evolution equations. The general time evolution of the classical ensemble can also account for the phenomenon of decoherence, corresponding to decreasing purity, and ``syncoherence'' for the increase of purity as the pure state fixed point is approached. This shows that the classical ensemble can describe an incompletely isolated quantum system embedded in its environment, with quantum mechanics as an idealization where the isolation becomes perfect.

In our picture, an atom and its environment are described by a classical statistical ensemble with infinitely many degrees of freedom. If a gas of atoms is dilute enough the picture of an isolated atom becomes a good approximation. Such an isolated atom can be described by a few observables out of the infinitely many possible observables of the whole system. The expectation values and correlations of these observables can be computed by a reduction to effective states, with ``effective probabilities'' mirrored in the density matrix. The limit of perfect isolation is described by a unitary time evolution - this is quantum mechanics.



\begin{thebibliography}{100}
\bibitem{EPR}A. Einstein, B. Podolski, N. Rosen, Phys. Rev. {\bf 47} (1935) 777
\bibitem{Bell}J. S. Bell, Physica 1 (1964) 195
\bibitem{3}C. Wetterich, in ``Decoherence and Entropy in Complex Systems'', ed. T. Elze, p. 180, Springer Verlag 2004, arXiv: quant-ph/0212031
\bibitem{CHSH}J. Clauser, M. Horne, A. Shimony, R. Holt, Phys. Rev. Lett. {\bf 23} (1969) 880
\bibitem{B71}J. Bell, ``Foundations of Quantum Mechanics'', ed. B. d'Espagnat, (New York: Academic, 1971) p. 171
\bibitem{CH}J. Clauser, M. Horne, Phys. Rev. {\bf D10} (1974) 526
\bibitem{CS}J. Clauser, A. Shimony, Rep. Prog. Phys. {\bf 41} (1978) 1881
\bibitem{POB}A. S. Holevo, ``Probabilistic and Statistical Aspects of Quantum Theory'' (Amsterdam, North Holland) 1982;\\
S. T. Ali, E. Prugovecki, J. Math. Phys. {\bf 18} (1977) 219;\\
M. Singer, W. Stulpe, J. Math. Phys. {\bf 33} (1992) 131
\bibitem{BB}E. Beltrametti, S. Bugajski, J. Phys. A: Math. Gen. {\bf 28} (1995) 3329; Int.~J.~Theor.~Phys. {\bf 34} (1995) 1221; \\
S.~Bugajski, Int.~J.~Theor.~Phys. {\bf 35} (1996) 2229;
\bibitem{Neu}G. Birkhoff, J. von Neumann, The Logic of Quantum mechanics, vol. {\bf 37} (1936); J.~von~Neumann, Mathematical Foundations of Quantum Mechanics, Princeton University Press (1955)
\bibitem{SB}W. Stulpe, P. Busch, J. Math. Phys. {\bf 49} (2008), 3
\bibitem{MIS}B. Misra ``Physical Reality and Mathematical Description'', p. 455 eds. C.~P.~Enz, J.~Mehra (Dordrecht, Reidel) (1974)
\bibitem{KS}S. Kochen, E. P. Specker, Journal of Mathematics and Mechanics {\bf 17} (1967), 59;\\
N. D. Mermin, Phys. Rev. Lett. {\bf 65} (1990) 3373;\\
A. Peres, J. Phys. A: Math. Gen. {\bf 24} (1991) L175\\
N. Straumann, arXiv: 0801.4931 [quant-ph]
\bibitem{CWN}C. Wetterich, Journal of Phys. 174 (2009) 012008, arXiv: 0811.0927[quant-ph]
\bibitem{14A}C. Wetterich, arXiv: 0906.4919 [quant-ph]
\bibitem{N}C. Wetterich, arXiv: 0809.2671 [quant-ph]
\bibitem{DC}H. D. Zeh, Found. Phys. {\bf 1} (1970) 69;\\
E. Joos, H. D. Zeh, Z. Phys. {\bf B59} (1985) 273;\\
E. Joos, H. D. Zeh, C. Kiefer, D. Giulini, J. Kupsch, I.-O. Stamatescu,
``Decoherence and the appearance of the classical world'', Springer 2003; \\
W. Zurek, Rev. Mod. Phys. {\bf 75} (2003) 715
\bibitem{Zu}W. Zurek, arXiv: 0707.2832
\bibitem{Zo}J. I. Cirac, P. Zoller, Phys. Rev. Lett. {\bf 74} (1995) 4091
\bibitem{Ze}D. Bouwmeester, J. W. Pan, K. Mattle, M. Eibl, H. Weinfurter, A. Zeilinger, Nature {\bf 390} (1997) 575
\bibitem{19A}C. Wetterich, arXiv: 0904.3048 [quant-ph]; arXiv: 0911.1261 [quant-ph]
\bibitem{GenStat}C. Wetterich, Nucl. Phys. {\bf B314} (1989) 40; Nucl. Phys. {\bf B397} (1993) 299
\bibitem{MP}G. C. Ghirardi, A. Rimini, T. Weber, Nuovo Cimento {\bf 36} (1976) 97;\\
R. L. Hudson, G. R. Moody, Z. Wahrscheinlichkeitstheorie verw. Gebiete {\bf 33} (1976) 343;\\
R. L. Hudson, Found. Phys. {\bf 11} (1981) 805
\end{thebibliography}
\end{document}